\pdfoutput=1
\RequirePackage{ifpdf}
\ifpdf
\documentclass[pdftex]{sigma}
\else
\documentclass{sigma}
\fi

\usepackage{esint}

\DeclareMathOperator{\Ker}{Ker}

\DeclareMathOperator{\im}{Im}
\DeclareMathOperator{\Hull}{Hull}
\DeclareMathOperator{\cls}{cls}
\DeclareMathOperator{\Span}{Span}
\newtheorem{ques}{Question}

\begin{document}

\allowdisplaybreaks

\renewcommand{\thefootnote}{$\star$}

\renewcommand{\PaperNumber}{020}

\FirstPageHeading

\ShortArticleName{The Sturm--Liouville Hierarchy of Evolution Equations}

\ArticleName{The Sturm--Liouville Hierarchy of Evolution Equations\\
and Limits of Algebro-Geometric Initial Data\footnote{This paper is a~contribution to the Special Issue in honor of
Anatol Kirillov and Tetsuji Miwa.
The full collection is available at \href{http://www.emis.de/journals/SIGMA/InfiniteAnalysis2013.html}
{http://www.emis.de/journals/SIGMA/InfiniteAnalysis2013.html}}}

\Author{Russell JOHNSON~$^\dag$ and Luca ZAMPOGNI~$^\ddag$}

\AuthorNameForHeading{R.~Johnson and L.~Zampogni}

\Address{$^\dag$~Dipartimento di Sistemi e Informatica, Universit\`{a} di Firenze, Italy}
\EmailD{\href{mailto:johnson@dsi.unifi.it}{johnson@dsi.unifi.it}}

\Address{$^\ddag$~Dipartimento di Matematica e Informatica, Universit\`{a} degli Studi di Perugia, Italy}
\EmailD{\href{mailto:zampoglu@dmi.unipg.it}{zampoglu@dmi.unipg.it}}

\ArticleDates{Received October 17, 2013, in f\/inal form February 27, 2014; Published online March 05, 2014}

\Abstract{The Sturm--Liouville hierarchy of evolution equations was introduced in~[\textit{Adv. Nonlinear Stud.} \textbf{11} (2011), 555--591] and includes the Korteweg--de Vries and the Camassa--Holm hierarchies.
We discuss some solutions of this hierarchy which are obtained as limits of algebro-geometric solutions.
The initial data of our solutions are (generalized) re\-f\/lec\-tion\-less Sturm--Liouville potentials~[\textit{Stoch. Dyn.} \textbf{8} (2008),   413--449].}


\Classification{37B55; 35Q53; 34A55; 34B24}

\renewcommand{\thefootnote}{\arabic{footnote}}
\setcounter{footnote}{0}

\section{Introduction}

It is well-known that the Korteweg--de Vries equation
\begin{gather}
\frac{\partial u}{\partial t}=6u\frac{\partial u}{\partial x}-\frac{\partial^3 u}{\partial x^3},
\qquad
u=u(t,x),\nonumber
\\
u(0,x)=u_0(x)\label{01}
\end{gather}
can be solved for various classes of initial data $u_0(\cdot)$ by making systematic use of the fact that it is formally
equivalent to the Lax equation
\begin{gather*}
\frac{dL_t}{dt}=[P,L_t],
\end{gather*}
where $L_t$ is the Schr\"odinger operator
\begin{gather}
\label{02}
L_t=-\frac{d^2}{dx^2}+u(t,x),
\end{gather}
and $P$ is the antisymmetric operator
\begin{gather*}
P=-4\left[D^3-\frac{3}{4}\left(Du+uD\right)\right],
\qquad D=\frac{d}
{dx}.
\end{gather*}
While this observation does not in and of itself provide a~solution of~\eqref{01}, it does imply that, if $u(t,x)$ is
a~decent solution of~\eqref{01}, then the spectrum of the operator $L_t$ in $L^2(\mathbb R)$ does not depend on $t$.

There are several interesting sets of initial data $\{u_0\}$ for which the Lax equation and the isospectral property of
the family $\{L_t\}$ can be used to solve equation~\eqref{01}.
Among these are the set of rapidly decreasing potentials~\cite{GGKM,Ma}, which contains in particular the the class of
classical re\-f\/lec\-tion\-less potentials~\cite{Cr,GGKM,JZ4,Ma}.
The latter class gives rise to the soliton solutions of the KdV equation.
Another family of initial data for which the Lax method ``works'' is that of the algebro-geometric
potentials~\cite{DMN,Mk1}.
The algebro-geometric potentials are quasi-periodic in~$x$.
By passing to appropriate limits, one can solve the KdV equation for more general almost periodic initial data;
see~\cite{Ego1,Ego,Le,Le1} for more information concerning this matter.

In 1985, Lundina~\cite{Lu} introduced the family $\operatorname{GR}$ of generalized re\-f\/lec\-tion\-less Schr\"odinger potentials, which
includes both the classical re\-f\/lec\-tion\-less potentials and (suitable translations of) the algebro-geometric potentials.
In succeeding years, it was shown that~\eqref{01} can be solved for various functions $u_0$ in $\operatorname{GR}$ (see,
e.g.,~\cite{GWZ,Le,Ma,Ma1,Mar}; also~\cite{Bel,GH1,NMPZ}).
In 2008, Kotani~\cite{Ko} proved that every element $u_0\in \operatorname{GR}$ gives rise to a~solution of~\eqref{01}, and indeed of
the entire KdV hierarchy of evolution equations.
He used the Sato--Segal--Wilson theory of the KdV hierar\-chy~\cite{Sa,SW}.
In fact, he was able to show that $\operatorname{GR}$ is contained in the Sato--Segal--Wilson family of potentials (see also~\cite{J86}
in this regard).
In~\cite{JZ4}, it was shown that if a~Sato--Segal--Wilson potential is suitably translated, then it lies in $\operatorname{GR}$.

It is also well-known that one can determine soliton solutions and algebro-geometric solutions for various other
nonlinear evolution equations and corresponding hierarchies, e.g., the Sine-Gordon equation and the nonlinear
Schr\"odinger equation.
We will not dwell on this matter here, but will only note that the Camassa--Holm equation~\cite{CH}
\begin{gather*}
y=2f-\frac{1}{2}\frac{\partial^2 f}{\partial x^2},
\\
\frac{\partial y}{\partial t}=\frac{\partial y}{\partial x}f+2y\frac{\partial f}{\partial x}
\end{gather*}
is related to the Sturm--Liouville operator def\/ined by
\begin{gather*}
-\varphi''+\varphi=\lambda y(x)\varphi
\end{gather*}
in a~fashion which is similar to the relation between the KdV equation~\eqref{01} and the Schr\"odinger
operator~\eqref{02} \cite{AF,BSS,GH,Z1}.
Motivated by this fact, we introduced in~\cite{JZ5,JZ6} a~hierarchy of evolution equations based on the general
Sturm--Liouville spectral problem
\begin{gather}\label{03}
-(p\varphi')'+q\varphi=\lambda y\varphi
\end{gather}
with positive weight $y$.
This so-called Sturm--Liouville hierarchy includes both the KdV and the Camassa--Holm hierarchies as well as other
evolution equations of interest (see Section~\ref{section3}).
We also worked out a~theory of algebro-geometric ``potentials'' $a=(p,q,y)$ for~\eqref{03} (see~\cite{JZ1,JZ2}), and
showed how one can produce the solutions of the various equations in the Sturm--Liouville hierarchy which admit a~given
algebro-geometric potential as an initial condition~\cite{JZ5}.

Now, one can also def\/ine the concept of ``generalized re\-f\/lec\-tion\-less Sturm--Liouville potentials'' (see~\cite{JZ3} and
Section~\ref{section2}).
However, there is as yet no analogue of the Sato--Segal--Wilson theory for the Sturm--Liouville potentials and the
Sturm--Liouville hierarchy.
For this and other reasons, it is of interest to construct solutions of the Sturm--Liouville hierarchy which have
initial values in the class $\operatorname{GRSL}$ of generalized re\-f\/lec\-tion\-less Sturm--Liouville potentials but are not of
algebro-geometric type.
Our goal in this paper is to make a~contribution in this direction.
We will in fact consider certain limits of algebro-geometric potentials, and construct the corresponding solutions of
the equations in the Sturm--Liouville hierarchy.

It is time to discuss in more detail the contents of the present paper.
Let $p$, $q$ and $y$ be real valued functions of $x\in\mathbb R$ such that: $p$, $q$ and $y$ are all bounded uniformly
continuous functions; $p\in C^1(\mathbb R)$ and has a~bounded uniformly continuous derivative $p'(x)$; $p$ and $y$
assume positive values and are bounded away from zero.
The dif\/ferential expression
\begin{gather*}
L_a=\frac{1}{y}\left\{-DpD+q\right\}
\qquad D=\frac{d}{dx}
\end{gather*}
def\/ines a~self-adjoint operator on the weighted space $L^2(\mathbb R,y(x)dx)$.
Suppose that this operator has spectrum
$\Sigma=[\lambda_0,\lambda_1]\cup[\lambda_2,\lambda_3]\cup\cdots\cup[\lambda_{2g},\infty)$.
The hypothesis that $a\in \operatorname{GRSL}$ ensures that $a$ is of algebro-geometric type, in the sense that information about
$a=(p,q,y)$ can be obtained by introducing the hyperelliptic Riemann surface $\mathcal R$ determined by the relation
$w^2=-(\lambda-\lambda_0)(\lambda-\lambda_1)\cdots(\lambda-\lambda_{2g})$, studying the motion of the zeroes of the
diagonal Green's function by using the holomorphic dif\/ferentials on $\mathcal R$ and the Abel map, etc.
These matters are discussed in~\cite{JZ1,JZ2,JZ3}, and part (but not all) of the discussion there is parallel to that
found in previous literature on algebro-geometric solutions of hierarchies of evolution equations.

Let now suppose that the f\/inite sequence $\lambda_0<\lambda_1<\cdots< \lambda_{2g}$ is replaced by an inf\/inite sequence
$\lambda_0<\lambda_1<\cdots<\lambda_{2g}<\cdots$ which tends to a~limit $\lambda_\ast\leq\infty$.
Set $\Sigma=\displaystyle{\bigcup_{i=0}^\infty[\lambda_{2i},\lambda_{2i+1}]}$ if $\lambda_\ast=\infty$ and
$\Sigma=\displaystyle{\bigcup_{i=0}^\infty[\lambda_{2i},\lambda_{2i+1}]\cup [\lambda_\ast,\infty)}$ if
$\lambda_\ast<\infty$.
Let $a$ be a~generalized re\-f\/lec\-tion\-less Sturm--Liouville potential which has spectrum $\Sigma$.
It turns out that, under fairly general conditions on the sequence $\{\lambda_i\}$, such potentials exist, and moreover
they serve as initial conditions giving rise to solutions of the Sturm--Liouville hierarchy.
These facts were proved in~\cite{JZ6} when $\lambda_\ast=\infty$, and our goal in the present paper is to prove them
when $\lambda_\ast<\infty$.
In particular we will obtain solutions of the Camassa--Holm hierarchy with generalized re\-f\/lec\-tion\-less initial data
which, so far as we know, are new.

The proof of the existence of a~generalized re\-f\/lec\-tion\-less Sturm--Liouville potential with spectrum $\Sigma$ proceeds by
algebro-geometric approximation, as does the proof of the existence of a~corresponding solution of the Sturm--Liouville
hierarchy.
This technique has been applied in the KdV case by several authors~\cite{BM,Ego1,Ego,Le,Le1,Mar,Z2}.
In the present case we f\/ind it convenient to deal with certain inf\/inite products by using convergence factors similar to
those of Weierstrass--Runge in the classical approximation theory of meromorphic functions~\cite{Kn}.
So far as we know, this method has not been used when working out solutions of hierarchies of evolution equations by
algebro-geometric approximation.
We will see that it is quite convenient in the case of the Sturm--Liouville hierarchy.

\looseness=-1
The paper is organized as follows.
In Section~\ref{section2} we recall some basic facts concerning the algebro-geometric Sturm--Liouville potentials~\cite{JZ1}.
In Section~\ref{section3} we review the construction of the Sturm--Liouville hierarchy of evolution equations and its solution for
algebro-geometric initial data.
We introduce the Weierstrass--Runge convergence factors~\cite{Kn} which, although unimportant in the algebro-geometric
setting, seem necessary in order to manage the potentials and solutions which arise as limits when $g\rightarrow\infty$
and $\lambda_{2g}\rightarrow\lambda_\ast<\infty$.
Finally, in Section~\ref{section4} we present the main results of this paper.
Namely, we construct solutions of the Sturm--Liouville hierarchy whose initial data $a=(p,q,y)$ are of generalized
re\-f\/lec\-tion\-less type, for which the corresponding operator has spectrum
$\Sigma=\displaystyle{\bigcup_{i=0}^\infty[\lambda_{2i},\lambda_{2i+1}]\cup[\lambda_\ast,\infty)}$ with
$\lambda_0<\lambda_1<\cdots<\lambda_{2g}<\cdots\rightarrow\lambda_\ast<\infty$.

\section{Some results on the Inverse Sturm--Liouville problem}\label{section2}
In this Section, we review some material concerning the study of the spectral theory of the Sturm--Liouville operator.
For a~detailed discussion concerning this topic, the reader is referred to~\cite{JZ1,JZ2,JZ3}.

Let $\mathcal E_2=\{b=(p,\mathcal M):\mathbb R\rightarrow\mathbb R^2 \,|\, b$ is uniformly continuous and $p(x)\geq
\delta, $ $\delta\leq \mathcal M(x)\leq \Delta$ for every $x\in\mathbb R\}$.
Further, let $\mathcal E_3=\{a=(p,q,y):\mathbb R\rightarrow\mathbb R^3\,|\,a$ is uniformly continuous and bounded,
$p(x)\geq \delta$, $\delta\leq y(x)\leq \Delta$ for every $x\in\mathbb R\}$.
Equip both $\mathcal E_2$ and $\mathcal E_3$ with the standard topology of uniform convergence on compact subsets of
$\mathbb R$.
Denote by $D$ the operator of dif\/ferentiation with respect to $x$.
If $a\in\mathcal E_3$ the Sturm--Liouville operator
\begin{gather*}
L_a:\mathcal D\rightarrow L^2(\mathbb R,ydx): \varphi\mapsto \frac{1}{y}(-DpD+q)\varphi
\end{gather*}
is def\/ined in its domain $\mathcal D=\{\varphi:\mathbb R\rightarrow\mathbb R\,|\,\varphi\in L^2(\mathbb R,y(x)dx)$,
$\varphi'$ is absolutely continuous and $\varphi''\in L^2(\mathbb R,y(x)dx)\}$.
With a~slight abuse of terminology, we refer to an element $a\in\mathcal E_3$ as a~{\it potential.}

Now, $L_a$ admits a~self-adjoint extension to all $L^2(\mathbb R,y(x)dx)$ (and we will continue to denote by $L_a$ this
extension as well), hence its spectrum $\Sigma_a$ is contained in $\mathbb R$, is bounded below and unbounded above, and
its resolvent set $R_a=\mathbb R
\setminus \Sigma_a$ is at most a~countable union (possibly unbounded) of disjoint open intervals.
Notice that the operators we are dealing with include the Schr\"odinger operator (obtained with $a=(1,q,1)$) and the
so-called acoustic operator (when $a=(1,1,y)$).

As already remarked in the Introduction, the spectral theory of the Sturm--Liouville operator is important both for its
intrinsic value, and for the connection existing between this kind of operator and the solutions of some important
evolution equations such as the KdV equation, the Camassa--Holm equation, and other recently discovered evolution
equations.

So, we discuss some facts concerning the spectral theory of the Sturm--Liouville operator.
It has turned out that it is convenient to attack this problem by using instruments of the theory of nonautonomous
dynamical systems.
To each $a\in\mathcal E_3$ and the corresponding operator $L_a$, one associates the eigenvalue equation
\begin{gather*}
E_a(\varphi,\lambda):=-(p\varphi')'+q\varphi=\lambda y\varphi,
\qquad \lambda\in\mathbb C .
\end{gather*}
This equation can be expressed as follows
\begin{gather*}
X'=A(x,\lambda)X=
\begin{pmatrix}
0&1/p(x)
\\
q(x)-\lambda y(x)&0
\end{pmatrix}
X,
\qquad X=
\begin{pmatrix}
\varphi(x)
\\
p(x)\varphi'(x)
\end{pmatrix}.
\end{gather*}

Now, let $A:\mathcal E_3\times\mathbb C\rightarrow\mathbb M(2,\mathbb C):(a,\lambda)\mapsto A(0,\lambda)$.
Denote by $\{\tau_s\}$ the Bebutov (or translation) f\/low on $\mathcal E_3$, i.e., if $a(\cdot)\in\mathcal E_3$, we
def\/ine $\tau_s(a)=a(s+\cdot)\in\mathcal E_3$.
Fix $a_0\in\mathcal E_3$, and let $\mathcal A=\cls\{\tau_s(a_0)\,|\,s\in\mathbb R\}$ ($\cls$ denotes the topological
closure).
One calls $\mathcal A$ the {\it Hull} of $a_0$ and writes $\mathcal A=\Hull(a_0)$.
Since $a_0$ is uniformly continuous, then $\mathcal A$ is a~compact subset of $\mathcal E_3$.
Moreover $\mathcal A$ is also invariant, in the sense that $\tau_s(\mathcal A)=\mathcal A$ for every $s\in\mathbb R$.
This construction (said to be of Bebutov type) allows one to use the instruments of topological dynamics to study the
spectral theory of the operators.
We will not pause to show how this takes place, however, we will brief\/ly introduce some objects which will be important
in the following pages.

It is clear that the construction we made above leaves us with a~family of linear systems, namely
\begin{gather}
\label{31}
\begin{pmatrix}
\varphi
\\
p\varphi'
\end{pmatrix}
'=A(\tau_x(a),\lambda)
\begin{pmatrix}
\varphi
\\
p\varphi'
\end{pmatrix},
\qquad
a\in\mathcal A,
\qquad
\lambda\in\mathbb C.
\end{gather}
The fundamental tool to study the systems~\eqref{31} is the concept of {\it exponential dichotomy.}
For $a\in\mathcal A$ and $\lambda\in\mathbb C$, let $\Phi_a(x)$ be the fundamental matrix solution of the corresponding
equation in~\eqref{31}:

\begin{definition}
The family~\eqref{31} is said to have an \textit{exponential dichotomy} over $\mathcal A$ if there are positive
constants $\eta$, $\rho$, together with a~continuous, projection valued function $P:\mathcal A\rightarrow\mathbb
M_2(\mathbb C)$ such that the following estimates holds:
\begin{itemize}
\item[(i)] $|\Phi_a(x)P(a)\Phi_a(s)^{-1}|\leq \eta e^{-\rho(x-s)},
\qquad x\geq s$, \item[(ii)] $|\Phi_a(x)(I-P(a))\Phi_a(s)^{-1}
|\leq \eta e^{\rho(x-s)},
\qquad x\leq s$.
\end{itemize}
\end{definition}

One has the following fundamental result (see~\cite{J86b,JM}).
\begin{theorem}
Let $\mathcal A$ be obtained by a~Bebutov type construction as above.
Consider the family~\eqref{31}.
If $a\in\mathcal A$ has dense orbit, then the spectrum $\Sigma_a$ of the operator $L_a$ equals the set
\begin{gather*}
\Sigma_{\rm ed}:=\{\lambda\in\mathbb C\,|\, \text{the family~\eqref{31} does \textbf{not} admit an exponential dichotomy
over} \;\mathcal A\}.
\end{gather*}
\end{theorem}
It is known that, if $\Im\lambda\neq 0$, then the family~\eqref{31} admits an exponential dichotomy over $\mathcal A$
(and indeed $\Sigma_a
\subset\mathbb R$).
Moreover, if $a\in\mathcal E_3$ and $\mathcal A=\Hull(a)$ then the spectrum of $L_{a}
$ and that of all the operators $L_{\tau_x(a)}$ coincide, i.e., $\Sigma_a=\Sigma_{\tau_x(a)}=\Sigma_{\rm ed}$ for every
$x\in\mathbb R$~\cite{FJZ}.

Now, let $a\in\mathcal E_3$ and let us f\/ix the Dirichlet boundary condition $\varphi(0)=0$.
There are well-def\/ined unbounded self-adjoint operators $L_a^\pm$ which are def\/ined in $L^2(\mathbb R^\pm,y(x)dx)$ and
which are determined via the relation
\begin{gather*}
L_a(\varphi)=\frac{1} {y}[-(p\varphi')'+q\varphi]
\end{gather*}
and the Dirichlet boundary condition at $x=0$.
If $\Im\lambda\neq 0$, we def\/ine the Weyl $m$-functions $m_\pm(a,\lambda)$ as those complex numbers which parametrize
$\Ker P(a)$ and $\im P(a)$, as follows:
\begin{gather*}
\im P(a)= \Span
\begin{pmatrix}
1
\\
m_+(a,\lambda)
\end{pmatrix},
\qquad \Ker P(a)=\Span
\begin{pmatrix}
1\\
m_-(a,\lambda)
\end{pmatrix}.
\end{gather*}
Note that, since $a\in\mathcal A$ and $\det\Phi_a(x)=1$ for every $x\in\mathbb R$, both $\Ker P(a)$ and $\im P(a)$ are
complex lines in $\mathbb C^2$.
Since $\tau_x(a)\in\mathcal A$ for every $x\in\mathbb R$, the functions $m_\pm(\tau_x(a),\lambda):=m_\pm(x,\lambda)$ are
well def\/ined.
They satisfy the Riccati equation
\begin{gather}
\label{ric}
m'+\frac{1}{p}m^2=q-\lambda y,
\qquad
\Im\lambda\neq 0.
\end{gather}

Next, let $a=(p,q,y)\in\mathcal E_3$ be a~Sturm--Liouville potential.
Consider the (unbounded, self-adjoint) operator $L_a=\frac{1}{y}(-DpD+q)$ on $L^2(\mathbb R,y(x)dx)$ with domain
$\mathcal D$.
We will def\/ine the Green's function for the operator~$L_a$.
The Green's function $\mathcal G_a(x,s,\lambda)$ is the kernel of the resolvent operator $(L_a-\lambda I)^{-1}$ acting
on $L^2(\mathbb R,y(x)dx)$ ($\Im\lambda\neq 0$).
This means that, if one considers the nonhomogeneous equation $-(p\psi')'+q\psi=\lambda y\psi+yf$, where $f\in
L^2(\mathbb R,y(x)dx)$ and if $\Im\lambda\neq 0$, one has
\begin{gather*}
\psi(x)=\int_{\mathbb R}\mathcal G_a(x,s,\lambda)f(s)ds.
\end{gather*}
If $a\in\mathcal A$, the Weyl $m$-functions $m_\pm(x,\lambda)$ and the {\it diagonal Green's function} $\mathcal
G_a(x,\lambda):=\mathcal G_a(x,x,\lambda)$ are connected by the fundamental relation
\begin{gather*}
\mathcal G_a(x,\lambda)=\frac{y(x)}{m_-(x,\lambda)-m_+(x,\lambda)},
\qquad\Im\lambda\neq 0.
\end{gather*}
The above formula implies that
\begin{gather*}
\mathcal G_a(x,\lambda)=\mathcal G_{\tau_x(a)}
(0,0,\lambda),
\qquad x\in\mathbb R,
\qquad
\Im\lambda\neq 0.
\end{gather*}

It is known that, for every $x\in\mathbb R$, the non-tangential limit
\begin{gather*}
\mathcal G_a(x,\eta):=\lim_{\varepsilon\rightarrow 0}\mathcal G_a(x,\eta+i\varepsilon)
\end{gather*}
exists for a.a.\
$\eta\in\mathbb R$.
In general, it is the behavior of the function $\mathcal G_a(x,\lambda)$ which provides a~division of $\mathcal E_3$
into subsets which we will call {\it spectral classes.}
Here, we mention only two of the most important spectral classes which exist, namely the algebro-geometric and the
re\-f\/lec\-tion\-less spectral classes.
\begin{definition}
(I)
A potential $a\in\mathcal E_3$ belongs to the algebro-geometric spectral class (brief\/ly, is algebro-geometric) if
it enjoys the following properties:
\begin{enumerate}\itemsep=0pt
\item[1)]
the spectrum $\Sigma_a$ of the operator $L_a$ is a~f\/inite union of disjoint compact intervals, plus a~half-line:
\begin{gather*}
\Sigma_a=[\lambda_0,\lambda_1]\cup[\lambda_2,\lambda_3]\cup\dots\cup[\lambda_{2g},\infty).
\end{gather*}
\item[2)]
for every $x\in\mathbb R$, one has $\Re \mathcal G_a(x,\eta)=0$, for a.a.\
$\eta\in \Sigma_a$.
\end{enumerate}

(II)~A potential $a\in\mathcal E_3$ belongs to the re\-f\/lec\-tion\-less spectral class (or is simply re\-f\/lec\-tion\-less)~if:
\begin{enumerate}\itemsep=0pt
\item[1)]
the spectrum $\Sigma_a$ has locally positive Lebesgue measure, in the sense that if $\eta\in\Sigma_a$ and if $I
\subset \mathbb R$ is an open interval with $\eta\in I$, then $I\cap\Sigma_a$ has positive Lebesgue measure;
\item[2)]
for every $x\in\mathbb R$, there holds $\Re\mathcal G_a(x,\eta)=0$ for a.a.\
$\eta\in\Sigma_a$.
\end{enumerate}

(III)
A family of potentials $\{a\}
_{a\in\mathcal F}$ lies in the isospectral class of $a_0\in\mathcal E_3$ if, for every $a\in\mathcal F$, the spectrum of
the operator $L_a$ equals the spectrum of the operator $L_{a_0}$.
\end{definition}

It would perhaps be more appropriate to speak of generalized re\-f\/lec\-tion\-less instead of re\-f\/lec\-tion\-less potentials, but we
prefer the simpler terminology.
Our def\/inition follows Craig~\cite{Cr}.

The condition (2) in the above def\/initions has some fundamental consequences: indeed, it turns out that, for every
$x\in\mathbb R$, both the maps $\lambda\mapsto m_\pm(x,\lambda)$ ($\Im\lambda\neq 0$) extend holomorphically through
every open set contained in the spectrum $\Sigma_a$.
If $h_\pm(x,\lambda)$ denote these extensions, we have
\begin{gather*}
h_+(x,\lambda)=
\begin{cases}
m_+(x,\lambda),&\Im\lambda>0,
\\
m_-(x,\lambda),&\Im\lambda<0
\end{cases}
\qquad\text{and}
\qquad h_-(x,\lambda)=
\begin{cases}
m_-(x,\lambda),&\Im\lambda>0,
\\
m_+(x,\lambda),&\Im\lambda<0.
\end{cases}
\end{gather*}

Other fundamental properties of an algebro-geometric potential $a\in\mathcal E_3$ can be summarized as follows:
\begin{enumerate}\itemsep=0pt
\item
The spectrum $\Sigma_a$ does not contain any isolated eigenvalues.
\item
The functions $m_\pm(a,\cdot)$ extend meromorphically through the resolvent set $R_a=\mathbb R
\setminus\Sigma_a$.
Let $I_j=[\lambda_{2j-1}
,\lambda_{2j}]$ be the closure of an interval of the resolvent set ($j=1,\dots,g$).
It turns out that in $I_j$ there exists exactly one point $P_j(a)$ with the following property: either
$m_+(a,P_j(a)+i\varepsilon)$ or $m_-(a,P_j(a)+i\varepsilon)$ has a~simple pole as $\varepsilon\rightarrow 0$.
The points $P_j(a)$ correspond to the isolated eigenvalues of the half-line restricted operators $L^\pm_a$ with the
boundary condition $\varphi(0)=0$.
\item
The properties (1) and (2) hold also for every potential $\tau_x(a)$ ($x\in\mathbb R$), hence we are left with the
functions $m_\pm(x,\lambda)$ which extend meromorphically through the resolvent set, and with the poles
$P_j(x):=P_j(\tau_x(a))$.
\end{enumerate}

The observations made so far have an important consequence.
Let $a\in\mathcal E_3$ be algebro-geometric, with spectrum
$\Sigma_a=[\lambda_0,\lambda_1]\cup[\lambda_2,\lambda_3],\cup\dots\cup[\lambda_{2g},\infty)$.
To such a~potential $a$ there are associated the poles $P_1(x),\dots,P_g(x)$ described in the above lines.
Let us assume from now on $\lambda_0>0$ -- though of course one can def\/ine and discuss algebro-geometric Sturm--Liouville
potentials when $\lambda_0<0$ (see~\cite{JM}).
  Let $\mathcal R$ be the Riemann surface of the relation
\begin{gather*}
w^2=-(\lambda-\lambda_0)(\lambda-\lambda_1)\cdots(\lambda-\lambda_{2g}).
\end{gather*}
Then $\mathcal R$ is a~torus with $g$ holes which correspond to the {\it spectral gaps}
$I_j=[\lambda_{2j-1},\lambda_{2j}]$ ($j=1,\dots,g$).
It is a~standard method now to consider the projection $\pi:\mathcal R\rightarrow \mathbb C_\infty$ (where $\mathbb
C_\infty$ is the Riemann sphere).
The projection $\pi$ is 2-1, except at the points $\lambda_0,\lambda_1,\dots,\lambda_{2g},\infty$ where it is 1-1.
We call these points the {\it ramification points} of $\mathcal R$.
If $\lambda\in\mathbb C_\infty$ is not a~ramif\/ication point, then there are two points $P_+$ and $P_-$ on $\mathcal R$
such that $\pi(P_\pm)=\lambda$.
Def\/ine a~function $k(P)$ on $\mathcal R$ by setting $k^2(\lambda)=-(\lambda-\lambda_0)\cdots(\lambda-\lambda_{2g})$, then
letting $k(0^\pm)$ be the positive/negative square root of $\lambda_0\lambda_1\cdots\lambda_{2g}$, and then extending via
analytic continuation.
The result is a~well-def\/ined function $P\mapsto k(P)$ on $\mathcal R$.

Further, let us def\/ine $c_j=\pi^{-1}([\lambda_{2j-1},\lambda_{2j}])$ ($j=1,\dots,g$).
Then $c_j$ are circles corresponding to the inner boundary of $\mathcal R$.
For a~point $P_j\in c_j$, its projection $\pi(P_j)$ lies in $I_j$.
We agree that~$k(P_j)$ is positive or negative according to the position of~$P_j$ on the circle~$c_j$.
In particular, if we express $P_j$ as
\begin{gather*}
P_j=(\lambda_{2j-1}-\lambda_{2j})
\sin^2\frac{\theta_j}
{2}+\lambda_{2j},
\qquad \theta_j\in[0,2\pi],
\end{gather*}
then $k(P_j)>0$ if $\theta_j\in(0,\pi)$, while $k(P_j)<0$ if $\theta_j\in(\pi,2\pi)$.
This convention will exclude every possible misunderstanding in the future.
Also, we will often commit an abuse of notation in denoting by $P_j$ both the point in $c_j$ and its projection in
$I_j$.

The setting we have introduced clarif\/ies the reason of the name {\it algebro-geometric.}
Indeed, the spectral properties of the operator $L_a$ can now be described by moving to the Riemann surface
$\mathcal R$.
We brief\/ly discuss this (see~\cite{JZ1,JZ2} for details).

Let $a=(p,q,y)\in\mathcal E_3$ be an algebro-geometric potential.
Hence its spectrum $\Sigma_a=[\lambda_0,\lambda_1]\cup\dots\cup[\lambda_{2g,\infty})$ is given, the Weyl $m$-functions
$m_\pm(x,\lambda)$ behave properly, together with their poles $P_1(x),\dots,P_g(x)$ (or, equivalently, the isolated
eigenvalues of the half-line restricted opera\-tors~$L^\pm_{\tau_x(a)}$).
If we concentrate further on the behavior of the Weyl $m$-functions, we can argue that, for every $x\in\mathbb R$, one
can def\/ine a~single meromorphic function $M:\mathcal R\rightarrow\mathbb C_\infty$ by setting, as before,
$M(x,0^\pm)=m_\pm(x,0)$, and then using analytic continuation on $\mathcal R$.
Again, for every f\/ixed $x\in\mathbb R$, one can def\/ine $m_+(x,P)=M(x,P)$ and $m_-(x,P)=M(x,
\sigma(P))$, where $\sigma$ is the hyperelliptic involution (the map which changes the sheets).
Actually, all these maps are jointly continuous when viewed as def\/ined on $\mathbb R\times(\mathbb C\setminus\mathbb
R)$.
Now, expanding $M$ at $\infty$, we obtain
\begin{gather*}
M(x,P)=m_+(x,P)=\frac{Q(x,\lambda)+\sqrt{p(x)y(x)}
k(P)}{H(x,\lambda)},
\\
m_-(x,P)=\frac{Q(x,\lambda)-
\sqrt{p(x)y(x)}
k(P)}{H(x,\lambda)},
\end{gather*}
where $\lambda=\pi(P)$, $Q(x,\lambda)$ is a~polynomial in $\lambda$ of degree $g$, and
\begin{gather*}
H(x,\lambda)=\prod\limits_{i=1}^g(\lambda-\pi(P_i(x))).
\end{gather*}
Moreover, it turns out that
\begin{gather*}
Q(x,\lambda)=\frac{p(x)}{2}
\sqrt{p(x)y(x)}
\left(\frac{1}{
\sqrt{p(x)y(x)}
}H(x,\lambda)\right)_x.
\end{gather*}
Here and throughout all the paper the subscripts $(\cdot)_s$ denote the (partial) derivative with respect to a~variable
$s$.

Recall that $\lambda_0>0$.
Set $\mathcal M(x)=m_-(x,0)-m_+(x,0)$.
Using the Riccati equation~\eqref{ric}, one can show that
\begin{gather}
\label{pole}
P_{i,x}(x)=\frac{-\mathcal M(x)k(P_i(x))\prod\limits_{i=1}^gP_i(x)}{p(x)k(0^+)\prod\limits_{j\neq i}(P_j(x)-P_i(x))},
\qquad i=1,\dots, g.
\end{gather}
The equations in~\eqref{pole} provide a~system of $g$ ODE's.
The induced f\/low is intended to take place on $\mathbb R$, hence we must take care of the value $k(P)$, according to the
observations we made above.
However, it is possible to pass to polar coordinates and write down a~system for the angular coordinate $\theta_i(x)$ of
each pole $P_i(x)$, avoiding any type of confusion.
Clearly, given an initial condition $P_1(0),\dots,P_g(0)$, the system~\eqref{pole} admits a~unique, globally def\/ined
solution, which we call the {\it pole motion.}
Once the pole motion is determined, we can write down the so-called {\it trace formulas} for the potential $a=(p,q,y)$, namely
\begin{gather}
\label{y}
y(x)=\frac{\mathcal M^2(x)\prod\limits_{i=1}^gP_i^2(x)}{4p(x)k^2(0^+)},
\\
\label{q}
q(x)=y(x)\left(\lambda_0+
\sum\limits_{i=1}
^g\lambda_{2i-1}+\lambda_{2i}-2P_i(x)\right)+\tilde q(x),
\end{gather}
where
\begin{gather*}
\tilde q(x)=-\left(\frac{(p(x)y(x))_x}{4y(x)}\right)_x+\left(\frac{(p(x)y(x))_x}{4y(x)}\right)^2.
\end{gather*}
It is a~recent discovery (to appear in a~forthcoming paper~\cite{JZ7}) that the function $\tilde q(x)$ plays a~crucial
role in a~development of a~theory of Gel'fand--Levitan--Marchenko type for the Sturm--Liouville operator.

We f\/inish this section by establishing the way to reconstruct an algebro-geometric potential $a=(p,q,y)$ from some given
spectral data.
Let us f\/ix $(p,\mathcal M)\in\mathcal E_2$.
Choose the spectral parameters, namely the ordered set
\begin{gather*}
\big\{0<\lambda_0<\lambda_1\leq P_1(0)\leq \lambda_2<\lambda_3\leq P_2(0)\leq \lambda_4<\dots<\lambda_{2g-1}\leq P_g(0)\leq\lambda_{2g}\big\}.
\end{gather*}
Let $P_1(x),\dots,P_g(x)$ be the solution of the system~\eqref{pole} with initial condition $P_1(0),\dots,P_g(0)$.
Finally, def\/ine $y(x)$ and $q(x)$ as to satisfy the relations~\eqref{y} and~\eqref{q}.
The triple $a=(p,q,y)\in\mathcal E_3$ thus def\/ined is an algebro-geometric potential whose spectrum $\Sigma_a$ is given
by $\Sigma_a=[\lambda_0,\lambda_1]\cup\dots\cup[\lambda_{2g},\infty)$.

\section{The Sturm--Liouville hierarchy\\ of evolution equations revisited}\label{section3}

The Sturm--Liouville hierarchy of evolution equations has been introduced and studied in detail in~\cite{JZ5,JZ6}.
In those papers, we determined certain solutions of the hierarchy: namely, the algebro-geometric solutions and some
types of solutions whose initial data are related to particular classes of re\-f\/lec\-tion\-less potentials.
In this paper, we extend the family of solutions we are able to describe by enlarging the class of admissible initial
conditions.
The initial conditions we introduce in the following lie in the re\-f\/lec\-tion\-less spectral class as well.
They are of a~type which generalizes the Schroedinger potentials considered in~\cite{BM,Ego1,Ego,Levitan}.
Namely, these initial Sturm--Liouville data have spectrum which clusters at f\/inite points of $\mathbb R$.
To include these potentials in the discussion, we will need to slightly modify the structure of the hierarchy.
At f\/irst sight, some quantities we will introduce soon will not be signif\/icant, but they will be fundamental when
a~limit procedure will be carried out.

But let us start by describing what we mean by Sturm--Liouville hierarchy of evolution equations.
For convenience, we will f\/irst choose the initial data, then def\/ine the evolution equations which will be solved.
Let $(p,\mathcal M)\in\mathcal E_2$ and choose arbitrarily the spectral parameters, i.e., the~set
\begin{gather*}
\Lambda_g=\{0<\lambda_0<\lambda_1\leq P_1(0)\leq \lambda_2<\lambda_3\leq P_2(0)\leq \lambda_4<\dots<\lambda_{2g-1}\leq
P_g(0)\leq \lambda_{2g}\}.
\end{gather*}
Then an algebro-geometric potential $a=(p,q,y)\in\mathcal E_3$, and the associated Sturm--Liouville operator $L_a$ with
prescribed spectrum $\Sigma_g=[\lambda_0,\lambda_1]\cup\dots\cup[\lambda_{2g},\infty)$ can be determined.

For $\lambda\in\mathbb C$, let us set
\begin{gather*}
E_n(\lambda)=\exp\left(\lambda+\frac{\lambda^2}{2}+\dots+\frac{\lambda^n}{n}\right).
\end{gather*}
Next, f\/ix a~point $\lambda_\ast\in\mathbb R^+
\setminus \Lambda_g$.
Def\/ine a~function
\begin{gather}
\label{UG}
U_g(x,\lambda)=\frac{-2p(x)k(0^+)}{\mathcal M(x)\prod\limits_{i=1}^gP_i(x)}
\prod\limits_{i=1}^g\frac{\lambda-P_i(x)}{\lambda_\ast-\lambda}E_i\left(\frac{\lambda_\ast-\lambda_{2i}}{\lambda_\ast-\lambda}\right).
\end{gather}
Clearly $U_g(x,\cdot)$ is def\/ined in the punctured complex plane $\mathbb C
\setminus\{\lambda_\ast\}
$ and has an essential singularity at $\lambda=\lambda_\ast$.
Further, def\/ine
\begin{gather}
\label{KG}
\tilde k^2_g(\lambda)=(\lambda_0-\lambda)\prod\limits_{i=1}^g\left(\frac{\lambda-\lambda_{2i}}{\lambda-\lambda_\ast}\right)
\left(\frac{\lambda-\lambda_{2i-1}}{\lambda-\lambda_\ast}\right)E_i^2\left(\frac{\lambda_\ast-\lambda_{2i}}{\lambda_\ast-\lambda}\right).
\end{gather}
It is clear that the function $\tilde k^2_g(\lambda)$ is strictly related to the function $k(P)$ def\/ined in the previous
section.

Let us observe that in~\cite{JZ5,JZ6}, we introduced the analogues of the functions $\tilde k_g$ and $U_g$ in which the
terms $E_i$ and $\frac{1}{\lambda_\ast-\lambda}$ are not present.
They are introduced here with an eye to the limit procedure which will be carried out in Section~\ref{section4}.
We state without giving all the details that the theory of~\cite{JZ5,JZ6} can be developed beginning with $U_g$ and
$\tilde k_g$ as given in~\eqref{UG} and~\eqref{KG}, as well as the simpler forms of $U_g$ and $\tilde k_g$
in~\cite{JZ5,JZ6} (which, to repeat, do not have the functions~$E_i$ and $\frac{1}{\lambda_\ast-\lambda}$).
We proceed to outline this (modif\/ied) theory.

Choose an integer $0\leq k \leq g-1$, and def\/ine two additional functions $T_g(x,\lambda)$ and $V_g(x,\lambda)$ in such
a~way that
\begin{gather}
\label{TG}
T_g(x,\lambda)=\frac{p(x)}{2\lambda^k}\left(\frac{U_g(x,\lambda)}{p(x)}\right)_x,
\end{gather}
and
\begin{gather}
\label{VG}
T_{g,x}(x,\lambda)+\frac{1}{\lambda^kp(x)}(q(x)-\lambda y(x))(V_g(x,\lambda)-U_g(x,\lambda))=0.
\end{gather}
Set
\begin{gather*}
B_g=
\begin{pmatrix}
-T_g&\lambda^{-k}U_g/p
\\
\lambda^{-k}(q-\lambda y)V_g&T_g
\end{pmatrix}
\end{gather*}
and, as usual
\begin{gather*}
A=
\begin{pmatrix}
0&1/p
\\
q-\lambda y&0
\end{pmatrix}
.
\end{gather*}
It can be shown (see~\cite{JZ5,JZ6}) that the so-called {\it stationary zero-curvature relation} holds, namely
\begin{gather*}
-B_{g,x}+[A,B_g]=0,
\end{gather*}
where $[A,B_g]=AB_g-B_gA$ is the commutator of $A$ and $B_g$.
Moreover, there holds
\begin{gather*}
\frac{d}{dx}\det B_g=0,
\end{gather*}
which translates into the fundamental relation
\begin{gather}
\label{det}
\frac{p^2}{4}\left[\left(\frac{U_g}{p}\right)_x\right]^2+\frac{1}{p}(q-\lambda y)U_gV_g=\tilde k_g^2(\lambda).
\end{gather}

Actually, more can be proved.
We state the following result; see~\cite{JZ5,JZ6}.
\begin{theorem}
If a~potential $a=(p,q,y)\in\mathcal E_3$ is algebro-geometric with spectral parameters
\begin{gather*}
\Lambda_g=\big\{0<\lambda_0<\lambda_1\leq P_1(0)\leq \lambda_2<\lambda_3\leq P_2(0)\leq \lambda_4<\dots<\lambda_{2g-1}\leq
P_g(0)\leq \lambda_{2g}\big\},
\end{gather*}
then there exist functions $U_g(x,\lambda),\tilde k_g^2(\lambda), T_g(x,\lambda)$ and $V_g(x,\lambda)$ as in the
relations~\eqref{UG}--\eqref{VG} respectively such that the zero curvature relation
$-B_{g,x}+[A,B_g]=0$ holds, together with the relation~\eqref{det}.

Conversely, let $a\in\mathcal E_3$, and suppose that the left endpoint of the spectrum of $L_a$ equals $\lambda_0>0$.
Let $\mathcal M(x)=m_-(x,0)-m_+(x,0)$.
Suppose that one can determine $U_g(x,\lambda)$ together with the corresponding quantities $\tilde k^2_g(\lambda)$,
$T_g(x,\lambda)$ and $V_g(x,\lambda)$ so that relations~\eqref{UG}--\eqref{VG} hold, and so
that the zero curvature relation $-B_ {g,x}+[A,B_g]=0$ and~\eqref{det} are valid.
Then $a$ is of algebro-geometric type.
\end{theorem}

We are now ready to introduce the Sturm--Liouville hierarchy of evolution equations.
It is here that the integer $k$ becomes signif\/icant.
We let a~parameter $t$ enter into play.
One obtains functions $a(t,x)=(p(t,x),q(t,x),y(t,x))$ and $\mathcal M(t)$ producing the poles $P_1(t,x),\dots,P_g(t,x)$,
and functions as in~\eqref{UG}--\eqref{VG} where the variable $t$ is present.
For instance, we will have a~function
\begin{gather*}
U_g(t,x,\lambda)=\frac{-2p(t,x)k(0^+)}{\mathcal M(t,x)
\prod\limits_{i=1}^gP_i(t,x)}\prod\limits_{i=1}^g\frac{\lambda-P_i(t,x)}{\lambda_\ast-\lambda}E_i
\left(\frac{\lambda_\ast-\lambda_{2i}}{\lambda_\ast-\lambda}\right),
\end{gather*}
and so on.
In this way one has matrices $B_g(t,x,\lambda)$ and $A(t,x,\lambda)$.
If we force $a(t,\cdot)$ to lie in the algebro-geometric isospectral class of $a(0,\cdot)$, then for every $t\in\mathbb
R$ one has the stationary zero-curvature relation
\begin{gather*}
-B_{g,x}(t,x,\lambda)+[A(t,x,\lambda),B_g(t,x,\lambda)]=0
\end{gather*}
together with the relation~\eqref{det}, which now expresses the invariance with respect to $t$ of its r.h.s.\ member as well.

However, we must still determine the time evolution of the functions we have introduced.
We do this as follows: f\/ix an integer $r$ such that $0\leq k\leq r< g$.
Introduce a~new matrix $B_r(t,x,\lambda)$ of the form
\begin{gather*}
B_r(t,x,\lambda)=
\begin{pmatrix}
-T_r(t,x,\lambda)&\lambda^{-k}\frac{U_r(t,x,\lambda)}{p(t,x)}
\\[2mm]
\lambda^{-k}(q(t,x)-\lambda y(t,x))V_r(t,x,\lambda)&T_r(t,x,\lambda)
\end{pmatrix},
\end{gather*}
where $U_r$ is a~polynomial of degree $r$ in $\lambda$ (whose coef\/f\/icients depend on $t$ and $x$), and
$T_r(t,x,\lambda)$ and $V_r(t,x,\lambda)$ are def\/ined via the relations
\begin{gather}
\label{TR}
\left(\frac{1}{p}\right)_t-\lambda^{-k}\left(\frac{U_r}{p}\right)_x+\frac{2}{p}T_r=0,
\\
\label{VR}
T_{r,x}+\frac{\lambda^{-k}}{p}(q-\lambda y)(V_r-U_r)=0.
\end{gather}

We pose the following basic question~\cite{JZ5,JZ6}
\begin{ques}
\label{ques}

Can $U_g(t,x,\lambda)$ and $U_r(t,x,\lambda)$ be chosen in such a~way that
\begin{gather}
-B_{g,x}+[A,B_g]=0, \nonumber
\\
A_t-B_{r,x}+[A,B_r]=0, \nonumber
\\
\frac{d}{dx}\det B_g=0,
\qquad
\text{and~\eqref{det} holds} \label{h}
\end{gather}
for all $(t,x)\in\mathbb R^2$ and all $\lambda\neq \lambda_\ast$?
\end{ques}
It is understood that $B_g$ satisf\/ies the conditions discussed above, and that $B_r$ satisf\/ies certain auxiliary
conditions which will be discussed in due course (see~\cite{JZ5}).

The second equation in the system~\eqref{h} is called the {\it zero-curvature relation,} and the system~\eqref{h}
determines the Sturm--Liouville evolution equation of order $r$ in a~way which we will explain in a~few lines.
Before doing so, we point out that the f\/irst and the third equations in~\eqref{h} \textit{force} the potentials
$a(t,x)=(p(t,x),q(t,x),y(t,x))$ to lie in the same isospectral class of $a(x):=a(0,x)=(p(0,x),q(0,x),y(0,x))$, i.e., if
we f\/ix $a(0,x)$ as initial data, the whole motion $t\mapsto a(t,x)$ will take place in its isospectral class.
To change the initial data means to change the matrix $B_g(0,x,\lambda)$ and the r.h.s.\ of~\eqref{det}! This
change will have an ef\/fect on $B_r$ as well because of the zero-curvature relation!

Before answering the above question, we explain how it translates into a~single evolution equation.
Let us set $\tilde U_r=U_r/p$.
It turns out that $\tilde U_r$ must satisfy the relation
\begin{gather}
2\lambda^k(q-\lambda y)_t+\frac{p_t}{p}(q-\lambda y)+\left(p\left(\frac{p_t}{p}\right)_x\right)_x
\nonumber
\\
\qquad
=2(p(q-\lambda y))_x\tilde U_r+4p(q-\lambda y)\tilde U_{r,x}-(p(p\tilde U_{r,x})_x)_x.
\label{ru}
\end{gather}
We make the fundamental ansatz that $\tilde U_r$ (and hence $U_r$) be a~polynomial of degree $r$ in $\lambda$, i.e.,
\begin{gather*}
\tilde U_r(t,x,\lambda)=\sum\limits_{j=0}^rf_j(t,x)\lambda^j.
\end{gather*}
If this is true, then the relation~\eqref{ru} provides $r+2$ recursion relations: it can be shown that, once we f\/ix
a~pair $(p(t,x),\mathcal M(t,x))\in\mathcal E_2$, then one of the coef\/f\/icients of $\tilde U_r$ is determined without
using these recursion relations, hence $r$ of the recursion relations will be used to f\/ind all the coef\/f\/icients
$f_j(t,x)$.
There remain 2 relations.
These 2 relations are compatibility conditions for~\eqref{h}, and translate into 2 evolution equations, one for the
function $q(t,x)$ and the other for the function $y(t,x)$.
It is this pair of equations which we call the Sturm--Liouville hierarchy of evolution equations.
In more detail, these 2 equations correspond to the formulas in~\eqref{ru} when we try to determine the coef\/f\/icients of
$\lambda^k$ and $\lambda^{k+1}$.
They give rise to relations of the type (here
$f_{-1}=f_{r+1}=0$)
\begin{gather}
q_t=\mathcal Q_r(t,x,f_{k-1},f_{k},q,q_x,q_{xx},\dots,y,y_x,y_{xx},p,p_x,p_{xx},\dots),
\nonumber
\\
y_t=\mathcal Y_r (t,x,f_k,f_{k+1},q,q_x,q_{xx},\dots,y,y_x,y_{xx},p,p_x,p_{xx},\dots).
\label{SLE}
\end{gather}

Question~\ref{ques} can now be formulated in the following convenient form
\begin{ques}
\label{ques2}
Does there exist a~polynomial $U_r(t,x,\lambda)$ of degree $r$ in $\lambda$ $($and satisfying certain auxiliary conditions$)$,
\begin{gather*}
U_r(t,x,\lambda)=
\sum\limits_{j=0}
^rp(t,x)f_j(t,x)\lambda^j
\end{gather*}
such that, if $\tilde U_r=U_r/p$ and $T_r(t,x,\lambda)$ and $V_r(t,x,\lambda)$ are defined as in~\eqref{TR}
and~\eqref{VR}, then the system~\eqref{h} admits a~unique solution, once the triple $a(0,x)=(p(0,x),q(0,x),y(0,x))$ is
a~given algebro-geometric potential?
\end{ques}

Before giving an answer to Question~\ref{ques2}, we give concrete examples of some evolution equations which can be obtained with this procedure.
Let $k=0$, and f\/ix $p(t,x)=y(t,x)=1$.
Then $\tilde U_r=U_r$ and~\eqref{ru} reads
\begin{gather*}
2q_t=2q_x U_r+4(q-\lambda) U_{r,x}- U_{r,xxx}.
\end{gather*}
This is the standard KdV hierarchy~\cite{DMN}.
For $r=1$, set $ U_1(t,x,\lambda)=f_1(t,x)\lambda+f_0(t,x)$.
Then
\begin{gather*}
f_{1,x}(t,x)=0,
\\
2q_x(t,x)f_1(t,x) -4f_{0,x}(t,x)=0,
\\
2q_t(t,x)=2q_x(t,x)f_0(t,x)+4q(t,x)f_{0,x}(t,x)-f_{0,xxx}(t,x).
\end{gather*}
If $f_1(t,x)=c_1$, we obtain $c_1q_x(t,x)= 2f_{0,x}(t,x)$, which implies $f_0(t,x)=\frac{c_1}{2}q(t,x)+c_2$.
Hence the last relation in the system above gives us
\begin{gather*}
q_t(t,x)=\frac{3}{2}c_1q(t,x)q_x(t,x)-\frac{c_1}{4}q_{xxx}(t,x)+c_2q_x(t,x),
\end{gather*}
which is a~generalized version of the classical KdV equation.
If $c_1=1$ and $c_2=0$, we obtain the classical KdV equation, i.e.,
\begin{gather*}
q_t(t,x)=\frac{3}{2}q(t,x)q_x(t,x)-\frac{1}{4}q_{xxx}(t,x).
\end{gather*}

As another example, let us assume that $k=1$ and let $p(t,x)=q(t,x)=1$ be f\/ixed.
Then~\eqref{ru} translates to
\begin{gather*}
2\lambda^2y_t(t,x)=2\lambda y_x(t,x) U_r(t,x,\lambda)-4(1-\lambda y(t,x)) U_{r,x}(t,x,\lambda)+ U_{r,xxx}(t,x,\lambda).
\end{gather*}
This is a~version of the Camassa--Holm hierarchy (another one can be obtained by setting $k=r$ as in~\cite{GH}).
If $r=1$, a~possible solution is given by
\begin{gather*}
f_0=c_1,
\qquad
c_1y(t,x)+c_2=2f_1(t,x)-\frac{1}{2}f_{1,xx}(t,x),
\\
y_t(t,x)=y_x(t,x)f_1(t,x)+2y(t,x)f_{1,x}(t,x).
\end{gather*}
This system is a~generalized version of the Camassa--Holm equation.
The classical Camassa--Holm equation is obtained by setting $c_1=1$ and $c_2=0$ (see~\cite{CH}).

Again, let us set $p(t,x)\equiv \varepsilon$, $y(t,x)\equiv 1$, $k=0$ and $r=1$.
Then $\tilde U_1=U_1/\varepsilon$, and the equation~\eqref{ru} translates to the system
\begin{gather*}
f_1=c_1,
\qquad
c_1q_x=2f_0,
\qquad
q_t=\frac{3}{2}c_1qq_x-\frac{c_1\varepsilon}{4}q_{xxx}+c_2q_x.
\end{gather*}
If $c_1=4$ and $c_2=0$, then the compatibility condition is given by
\begin{gather*}
q_t=6qq_x-\varepsilon q_{xxx},
\end{gather*}
which is a~well-known and important generalization of the KdV equation, used in \cite{LL1,LL2,LL3,Ven} in connection
with Burger's equation, which is indeed the limit as $\varepsilon\rightarrow 0$ of such a~KdV generalization.

Moreover, if $p(t,x)=1$, $q(t,x)\equiv\varepsilon$, $k=1$ and $g=1$, then the compatibility condition reads (for
suitably chosen constants $c_1$ and $c_2$)
\begin{gather*}
4\varepsilon u_{1,t}-u_{1,xxt}=12\varepsilon u_1u_{1,x}-u_1u_{1,xxx}-2u_{1,x}u_{1,xx}.
\end{gather*}
This equation is a~generalization of the CH equation.
Its limit (whenever it exists) as $\varepsilon\rightarrow0$ is the Hunter--Saxton equation
\begin{gather*}
u_{1,xxt}=u_1u_{1,xxx}+2u_{1,x}u_{1,xx}.
\end{gather*}
Note that the constants in all the above constructions can be chosen at will.

Before proceeding with the discussion, we wish to make another observation: the fact that both the KdV and the
Camassa--Holm hierarchies are included in our hierarchy is not surprising at all.
Indeed, they are strictly related as one can use a~Liouville transform to move from one hierarchy to the
other~\cite{JZ4bis,MK}.

The answer to Question~\ref{ques2} is af\/f\/irmative.
In more detail, at f\/irst we choose (at will!!) a~family $(p(t,x),\mathcal M(t,x))\in\mathcal E_2$.
We then construct a~polynomial $U_r$ in the following way: the coef\/f\/icients of $U_r$ are determined recursively via the
relation
\begin{gather}
\tilde U_{r,x}(\lambda)=\frac{\lambda^k}{p}\left[\frac{\mathcal M_t}{\mathcal
M}+p\left(\frac{1}{p}\right)_t\right]-\frac{\mathcal M_x}{\mathcal M}\tilde U_r(\lambda)
+\sum\limits_{i=0}^n\left[\frac{\lambda^k}{P_i^k}\tilde U_r(P_i)-\tilde U_r(\lambda)\right]\frac{\lambda P_{i,x}}{P_i(\lambda-P_i)},
\label{recur}
\end{gather}
where we omitted to write down explicitly the dependence of the functions with respect to $t$ and~$x$.
Note that $p$ and $\mathcal M$ are known functions of $(t,x)$, while the functions ({\it poles}) $P_i=P_i(t,x)$ remain
to be determined.

Once this is done, we determine the poles $P_1(t,x),\dots,P_g(t,x)$ by solving the system
\begin{gather}
P_{i,x}(t,x)=\frac{-\mathcal M(t,x)k_g(P_i(t,x))\prod\limits_{i=1}^gP_i(t,x)}{p(t,x)k_g(0^+)\prod\limits_{j\neq
i}(P_j(t,x)-P_i(t,x))}
\qquad
(\text{as in~\eqref{pole}}),
\nonumber
\\
P_{i,t}(t,x)=\frac{U_r(t,x,P_i(t,x))}{P_i^k(t,x)}P_{i,x}(t,x)
\label{poles}
\end{gather}
together with the initial condition $P_1(0,0)\in[\lambda_1,\lambda_2],\dots,P_g(0,0)\in[\lambda_{2g-1},\lambda_{2g}]$.
Then it turns out that the system~\eqref{poles} is consistent, and that the polynomial $U_r(t,x,\lambda)$ gives rise,
via the corresponding matrix $B_r$, to a~solution of~\eqref{h}.

Moreover, one can write down the trace formulas (analogous to those in~\eqref{y} and~\eqref{q}):
\begin{gather*}
y(t,x)=\frac{\mathcal M^2(t,x)\prod\limits_{i=1}^gP_i^2(t,x)}{4p(t,x)k^2(0^+)},
\\
q(t,x)=y(t,x)\left(\lambda_0+\sum\limits_{i=1}^g\lambda_{2i-1}+\lambda_{2i}-2P_i(t,x)\right)+\tilde q(t,x),
\end{gather*}
where
\begin{gather*}
\tilde q(t,x)=-\left(\frac{(p(t,x)y(t,x))_x}{4y(t,x)}\right)_x+\left(\frac{(p(t,x)y(t,x))_x}{4y(t,x)}\right)^2.
\end{gather*}
The functions $q(t,x)$ and $y(t,x)$ are the solutions of the evolution equations~\eqref{SLE}, hence we have solved the
Sturm--Liouville evolution equation of order $r$.
Notice that, since the maps $x\mapsto P_i(t,x)$ satisfy the f\/irst equation in the system~\eqref{poles}, the triple
$a(t,\cdot)=(p(t,\cdot),q(t,\cdot),y(t,\cdot))$ lies in the isospectral class of the algebro-geometric potential
$a(0,\cdot)$ for every $t\in\mathbb R$, hence the map $t\mapsto a(t,x)$ is a~curve in the isospectral class of $a(0,x)$
starting from $a(0,x)$!

Let us further repeat that these developments can be carried out both in the case when $U_g$ contains the factors $E_i$
and $\frac{1}{\lambda_\ast-\lambda}$ and in the case when these factors are not present.

\section{Some solutions of the Sturm--Liouville hierarchy}\label{section4}

All the machinery we have discussed in the previous section works well when we take as initial conditions potentials of
algebro-geometric type.
What happens if we change the initial condition? Clearly, we cannot choose an initial condition at will, because the
structure of the hierarchy has to remain consistent.
In particular, the structure of the function $U_g$ must be preserved in some sense.
An idea is that of considering as initial data some re\-f\/lec\-tion\-less Sturm--Liouville potentials whose spectra consist of
inf\/initely many intervals clustering at $\infty$.
This has been done in~\cite{JZ6}.
In this case $U_g$ translates to an entire function $U(t,x)$ with the inf\/initely many zeros
$P_1(t,x),\dots,P_g(t,x),\dots$.
It is important, however, that instead $U_r$ remain a~polynomial of degree $r$.
The reader can be addressed to~\cite{JZ5,JZ6} for a~detailed discussion of these topics.

The purpose of this section is that of enlarging the class of initial conditions for which the Sturm--Liouville
hierarchy can be solved, by including other re\-f\/lec\-tion\-less potentials whose spectra can cluster at a~f\/inite real point
$\lambda_\ast$.
The discussion of these new potentials will require the introduction of the factors $E_i$ and
$\frac{1}{\lambda_\ast-\lambda}$ seen in the def\/inition of $U_g$ and $\tilde k_g$ given in~\eqref{UG} and~\eqref{KG}
respectively.

Before introducing a~suitable hierarchy of evolution equations, or rather a~zero-curvature relation which determines
such a~hierarchy, we should explain how to construct re\-f\/lec\-tion\-less potentials with some prescribed properties of the
spectrum of the associated operator.
We will use a~procedure which we call of {\it algebro-geometric approximation.}
The construction we are going to illustrate is described in detail in~\cite{JZ6} in the case when $\lambda_\ast=\infty$.
Let us f\/ix a~sequence of positive real numbers
\begin{gather*}
\tilde\Lambda=\{\lambda_0<\lambda_1<\lambda_2<\cdots<\lambda_{2g}<\cdots\}.
\end{gather*}
Set $I_k=[\lambda_{2k-1},\lambda_{2k}]$, $h_{ij}=\operatorname{dist}(I_i,I_j)$, $d_j=\lambda_{2j}-\lambda_{2j-1}$ and
$h_{0k}=\lambda_{2k-1}-\lambda_0$.
We assume that the sequence $\tilde\Lambda$ satisf\/ies the following assumptions:
\begin{gather*}
 {\rm (H1)}  \quad {\lim_{i\rightarrow\infty}\lambda_i=\lambda_\ast}, \\
{\rm (H2)} \quad {\sum\limits_{j=1}^\infty d_j<\infty} ,\\
{\rm (H3)} \quad {\sup_{j\in\mathbb N}\sum\limits_{k\neq j}\frac{\sqrt{d_k}}{h_{jk}}<\infty}.
\end{gather*}

We will construct a~Sturm--Liouville potential $a(x)=(p(x),q(x),y(x))\in\mathcal E_3$ which is re\-f\/lec\-tion\-less and such
that the spectrum of the associated operator $L_a$ is given by
\begin{gather*}
\Sigma=[\lambda_0,\lambda_1]\cup[\lambda_2,\lambda_3]\cup\cdots\cup[\lambda_{2g},\lambda_{2g+1}]\cup\cdots\cup[\lambda_\ast,\infty).
\end{gather*}
Actually, the method we will describe below can be applied to prove the existence of a~re\-f\/lec\-tion\-less Sturm--Liouville
potential such that the spectrum of the associated Sturm--Liouville operator is given by
\begin{gather*}
\Sigma= {\bigcap_{g\in\mathbb N}\Sigma_g},
\end{gather*}
where
\begin{gather*}
\Sigma_g=[\lambda_0,\lambda_1]\cup\dots\cup[\lambda_{2g},\lambda_\ast]\cup[\overline\lambda,\infty),
\end{gather*}
and $\overline\lambda$ is any real number strictly greater than $\lambda_\ast$.
Also, this method can be applied when there is more than one cluster point in the sequence $\{\lambda_i\}$, and in fact
when there is an arbitrary f\/inite number of cluster points $\big\{\lambda_\ast^{(1)},\dots,\lambda_\ast^{(k)}\big\}$.

However, to keep the discussion clearer, we will only deal with the case when
\begin{gather*}
\Sigma=[\lambda_0,\lambda_1]\cup[\lambda_2,\lambda_3]\cup\dots\cup[\lambda_{2g},\lambda_{2g+1}]\cup\dots\cup[\lambda_\ast,\infty).
\end{gather*}
The procedure is inspired by the following important proposition~\cite{JM} (see also~\cite{JZ3,JZ5,JZ6}).

\begin{proposition}
Let $\{a_n\}=\{(p_n,q_n,y_n)\}\subset\mathcal E_3$ be a~sequence of potentials such that $a_n\rightarrow a=(p,q,y)\in\mathcal E_3$ uniformly on
compact subsets of $\mathbb R$.
Assume that $a_n$ is reflectionless and that $\Sigma_{a_{n+1}}\subset\Sigma_{a_n}$ for every $n\in\mathbb N$.
Assume further that the set $\Sigma=\displaystyle{\bigcap_{n\in\mathbb N}\Sigma_{a_n}}$ has locally positive Lebesgue measure.
Then $a$ is reflectionless and the spectrum $\Sigma_a$ of the operator $L_a$ equals the set $\Sigma$.
\end{proposition}

We will not prove this proposition.
It uses the Weyl decreasing disc construction and some additional reasoning concerning the spectral measures and the
spectra of the operators $L_{a_n}$.

Inspired by the above proposition, we f\/ix the f\/inite set $\tilde\Lambda_g\subset\tilde\Lambda$ given by
\begin{gather*}
\tilde\Lambda_g=\{\lambda_0,\lambda_1,\dots,\lambda_{2g}\},
\end{gather*}
then choose points $P_j(0)\!\in\![\lambda_{2j-1},\lambda_{2j}]$, $j=1,\dots,g$.
Moreover, let us f\/ix a~pair \mbox{$(p(x),\mathcal M(x))\!\in\!\mathcal E_2$}.
In correspondence with these choices, one can construct an algebro-geometric potential
$a_g=(p(x),q_g(x),y_g(x))\in\mathcal E_3$ such that the spectrum of the operator $L_{a_g}$ is given by
\begin{gather*}
\Sigma_g=[\lambda_0,\lambda_1]\cup\dots\cup[\lambda_{2g},\infty),
\end{gather*}
and such that the trace formulas~\eqref{y} and~\eqref{q} hold, together with the system~\eqref{pole}.
Now we let~$g$ vary over $\mathbb N$.
We obtain sequences $\{a_g\}=\{p,q_g,y_g\}\in\mathcal E_3$ of algebro-geometric potentials and corresponding poles
$\{P_j^{(g)}(x)\}$.
Next, we let $g\rightarrow\infty$.
It can be shown that the sequences $\{P_j^{(g)}(x)\}\rightarrow\{P_j(x)\}$ for every $x\in\mathbb R$, and that
$a_g\rightarrow a=(p,q,y)\in\mathcal E_3$ uniformly on compact subsets of $\mathbb R$ (this convergence, however is not
uniform on $\mathbb R$~\cite{JZ3,JZ6}).
One can show that the poles $P_j(x)$ satisfy the following system of inf\/initely many ODE's ($j\in\mathbb N$)
\begin{gather}
P_{j,x}(x)=\pm\frac{\mathcal M(x)}{\sqrt{\lambda_0}p(x)}\left(\prod\limits_{k\in\mathbb N}\frac{P_k(x)}{\sqrt{\lambda_{2k-1}\lambda_{2k}}}\right)
\sqrt{(\lambda_{2j}-P_j(x))(P_j(x)-\lambda_{2j-1})}
\nonumber
\\
\phantom{P_{j,x}(x)=}
\times\left(\prod\limits_{k\neq j}\frac{\sqrt{(\lambda_{2k-1}-P_j(x))(\lambda_{2k}-P_j(x))}}{P_k(x)-P_j(x)}\right)\sqrt{P_j(x)-\lambda_0}.
\label{poleinf}
\end{gather}
The sign $\pm$ in the equations~\eqref{poleinf} comes from the necessity to choose a~sign of the square root $
\sqrt{(\lambda_{2k-1}
-P_j(x))(\lambda_{2k}-P_j(x))}$.
This ambiguity, however, can be avoided by passing to suitable angular coordinates
$\theta_1(x),\dots,\theta_n(x),\dots$.
But this is not the place in which to discuss this matter.

Once we have determined the pole motion, we can write the trace formulas
\begin{gather*}
y(x)=\frac{\mathcal M^2(x)}{4p(x)\lambda_0}\prod\limits_{k\in\mathbb N}\frac{P_k^2(x)}{\lambda_{2k-1}\lambda_{2k}},
\\
q(x)=y(x)\left(\lambda_0+\sum\limits_{k\in\mathbb N}\lambda_{2k-1}+\lambda_{2k}-2P_k(x)\right)+\tilde q(x),
\end{gather*}
where
\begin{gather*}
\tilde q(x)=-\left(\frac{(p(x)y(x))_x}{4y(x)}\right)_x+\left(\frac{(p(x)y(x))_x}{4y(x)}\right)^2.
\end{gather*}
The assumptions (H1)--(H3) are used to show that the appropriate quantities are well def\/ined and converge properly.
See~\cite{JZ3,JZ6} for the above developments.

Now we move to the main question of interest in this paper, namely the solution of the Sturm--Liouville hierarchy for
certain non algebro-geometric re\-f\/lec\-tion\-less initial data.

First, we introduce a~zero-curvature relation which takes into account the structure of the potential $a$ obtained
above.
To do this, let the family $\{(p(t,x),\mathcal M(t,x))\}\in\mathcal E_2$ (indexed by $t\in\mathbb R$) be f\/ixed.
Choose the set $\tilde\Lambda$ as above and initial data
$P_1(t,0)\in[\lambda_1,\lambda_2],\dots,P_g(t,0)\in[\lambda_{2g-1},\lambda_{2g}],\dots$ in such a~way that they vary
smoothly with respect to $t\in\mathbb R$.
Let $P_1^{(g)}(t,x),\dots, P_g^{(g)}(t,x)$ be the solution of the system
\begin{gather*}
P^{(g)}_{i,x}(t,x)=\frac{-\mathcal M(t,x)k_g(P^{(g)}_i(t,x))
\prod\limits_{i=1}^gP_i^{(g)}(t,x)}{p(t,x)k_g(0^+)\prod\limits_{j\neq i}(P_j^{g)}(t,x)-P_i^{(g)}(t,x))}.
\end{gather*}

For every f\/ixed $t\in\mathbb R$, let us construct the sequence $\{a_g(t,x)=(p(t,x),q_g(t,x),y_g(t,x))\}
\subset\mathcal E_3$ as above, and let $a(t,x)=(p(t,x),q(t,x),y(t,x))$ be its limit in $\mathcal E_3$ (we emphasize that
the variable here is $x$, while $t$ is considered as a~parameter).
Let
\begin{gather*}
U_g(t,x,\lambda)=\frac{-2p(t,x)k_g(0^+)}{\mathcal M(t,x)\prod\limits_{i=1}^gP_i^{(g)}(t,x)}
\prod\limits_{i=1}^g\frac{\lambda-P_i^{(g)}(t,x)}{\lambda_\ast-\lambda}E_i\left(\frac{\lambda_\ast-\lambda_{2i}}{\lambda_\ast-\lambda}\right),
\end{gather*}
where
\begin{gather*}
E_n(\lambda)=\exp\left(\lambda+\frac{\lambda^2}{2}+\dots+\frac{\lambda^n}{n}\right).
\end{gather*}
The function $\lambda\mapsto U_g(t,x,\lambda)$ is def\/ined in the region $G=\mathbb C\setminus \{\lambda_\ast\}$.

We prove the following
\begin{theorem}
As $g\rightarrow\infty$ the functions $U_g(t,x,\lambda)$ converge to a~holomorphic function $U(t,x,\lambda)$, uniformly
on compact subsets of $G$.
This convergence is uniform also with respect to $(t,x)\in\mathbb R^2$.
\end{theorem}
\begin{proof}
For every f\/ixed $t\in\mathbb R$, the poles $P_j^{(g)}(t,x)$ converge pointwise to poles $P_j(t,x)$ as
$g\rightarrow\infty$, where $P_j(t,x)$ satisfy the relation~\eqref{poleinf} ($j\in\mathbb N$).
Each $P_j(t,x)$ lies in the corresponding interval $I_j=[\lambda_{2j-1},\lambda_{2j}]$ for every $(t,x)\in\mathbb R^2$.

Now, if $g\rightarrow\infty$, the pointwise limit of $U_g(t,x,\lambda)$ is given by the function
\begin{gather}
\label{Uinf}
U(t,x,\lambda)
=-\frac{2\sqrt{\lambda_0}p(t,x)}{\mathcal M(t,x)}\prod\limits_{k=1}^\infty\frac{\sqrt{\lambda_{2k-1}\lambda_{2k}}}{P_k(t,x)}
\prod\limits_{i=1}^\infty\frac{\lambda-P_i(t,x)}{\lambda_\ast-\lambda}E_i\left(\frac{\lambda_\ast-\lambda_{2i}}{\lambda_\ast-\lambda}\right).
\end{gather}
However, the expression~\eqref{Uinf} has only informal signif\/icance at the moment, because we do not know if it exists
(the inf\/inite products must converge properly!).

The inf\/inite product
\begin{gather*}
\prod\limits_{k=1}^\infty\frac{
\sqrt{\lambda_{2k-1}
\lambda_{2k}}}{P_k(t,x)}
\end{gather*}
is well def\/ined, because
\begin{gather*}
\prod\limits_{k=1}^\infty\frac{
\sqrt{\lambda_{2k-1}
\lambda_{2k}}}{P_k(t,x)}\leq \prod\limits_{k=1}^\infty\frac{\lambda_{2k}}{P_k(t,x)},
\end{gather*}
and the series
\begin{gather*}
\sum\limits_{k=1}^\infty\left|1-\frac{\lambda_{2k}}{P_k(t,x)}\right|\leq
\sum\limits_{k=1}^\infty\left|\frac{d_k}{h_{0k}}\right|
\end{gather*}
converges, using the assumption (H2), uniformly with respect to $(t,x)\in\mathbb R^2$.
Hence, the main problem lies in proving that the inf\/inite product
\begin{gather*}
\prod\limits_{i=1}^\infty\frac{\lambda-P_i(t,x)}{\lambda_\ast-\lambda}E_i\left(\frac{\lambda_\ast-\lambda_{2i}}{\lambda_\ast-\lambda}\right)
\end{gather*}
exists.
Observe that, if the factors $E_i$ and $\frac{1}{\lambda_\ast-\lambda}$ are absent, the convergence {\it does not hold.}
Our use of these factors is motivated by the classical theory of Weierstrass and Runge~\cite{Kn}.

To prove this, let $K
\subset G$ be a~compact subset.
We claim that the series
\begin{gather*}
\sum\limits_{i=1}
^\infty\left|1-\frac{\lambda-P_i(t,x)}{\lambda_\ast-\lambda}E_i\left(\frac{\lambda_\ast-\lambda_{2i}}{\lambda_\ast-\lambda}\right)\right|
\end{gather*}
converges uniformly with respect to $\lambda\in K$ and $(t,x)\in\mathbb R^2$.
By a~well-known result on inf\/inite products, this will imply that the inf\/inite product under consideration is well
def\/ined.
Let us rewrite
\begin{gather}
\sum\limits_{i=1}^\infty\left|1-\frac{\lambda-P_i(t,x)}{\lambda_\ast-\lambda}E_i
\left(\frac{\lambda_\ast-\lambda_{2i}}{\lambda_\ast-\lambda}\right)\right|
\leq\sum\limits_{i=1}^\infty\left|1-\frac{\lambda-\lambda_{2i}}{\lambda_\ast-\lambda}E_i
\left(\frac{\lambda_\ast-\lambda_{2i}}{\lambda_\ast-\lambda}\right)\right|
\nonumber
\\
\qquad
+\sum\limits_{i=1}^\infty\left|\frac{\lambda_{2i}-P_i(t,x)}{\lambda_\ast-\lambda}E_i
\left(\frac{\lambda_\ast-\lambda_{2i}}{\lambda_\ast-\lambda}\right)\right|.
\label{aus}
\end{gather}
Let $D=\operatorname{dist}(\mathbb C \setminus G,K)$.
If $\lambda\in K$, then
\begin{gather*}
\left|\frac{\lambda_\ast-\lambda_{2i}
}{\lambda_\ast-\lambda}\right|\leq \frac{|\lambda_\ast-\lambda_{2i}|}{D}.
\end{gather*}
Since $\lambda_{2i}\rightarrow\lambda_\ast$, for every $0<\varepsilon<1$, there holds{\samepage
\begin{gather*}
\left|\frac{\lambda_\ast-\lambda_{2i}}{\lambda_\ast-\lambda}\right|\leq\varepsilon
\end{gather*}
for suf\/f\/iciently large $i\in\mathbb N$.}

Now, the factors of the second summand in the r.h.s.\ of~\eqref{aus} can be estimated as follows:
\begin{gather*}
\left|\frac{\lambda_{2i}-P_i(t,x)}{\lambda_\ast-\lambda}E_i\left(\frac{\lambda_\ast-\lambda_{2i}}{\lambda_\ast-\lambda}\right)\right|\leq
\frac{d_i}{D}\exp\left(\varepsilon+\frac{\varepsilon}{2}+\cdots+\frac{\varepsilon^i}{i}\right)
\end{gather*}
for suf\/f\/iciently large $i\in\mathbb N$.
Since the series $
\sum\limits_{i=1}
^\infty \frac{\varepsilon^i}{i}=D_1<\infty$, the second summand in the r.h.s.\ of~\eqref{aus} is dominated by
the uniformly convergent series
\begin{gather*}
\frac{e^{D_1}}{D}
\sum\limits_{i=1}
^\infty d_i<\infty.
\end{gather*}

We now move our attention to the f\/irst summand in the r.h.s.\ of~\eqref{aus}.
To prove that it converges, we can argue as follows.
Let us set $z=\frac{\lambda_\ast-\lambda_{2i}}{\lambda_\ast-\lambda}$.
Then
\begin{gather*}
\frac{\lambda-\lambda_{2i}}{\lambda_\ast-\lambda}E_i\left(\frac{\lambda_\ast-\lambda_{2i}}{\lambda_\ast-\lambda}\right)=(z-1)E_i(z)=F_i(z).
\end{gather*}
We expand $F_i(z)$ at $z=0$, so that
\begin{gather*}
F_i(z)=1+
\sum\limits_{k=1}
^\infty a_kz^k.
\end{gather*}
Dif\/ferentiating $F_i(z)$ with respect to $z$, we obtain
\begin{gather*}
\sum\limits_{k=1}
^\infty ka_kz^{k-1}=-z^iE_i(z).
\end{gather*}
Also $E_i(z)$ can be expanded at $z=0$, being an exponential function.
Doing so, one observes that $a_1,\dots,a_i=0$.
Moreover, since the coef\/f\/icients of the expansion of $E_i(z)$ are all positive, we must have
\begin{gather*}
|a_k|=-a_k
\end{gather*}
for every $k>i$.
This implies that
\begin{gather*}
0=F_i(1)=1+
\sum\limits_{k=i+1}
^\infty a_k,
\end{gather*}
hence
\begin{gather*}
\sum\limits_{k=i+1}
^\infty|a_k|=1.
\end{gather*}
But now we have
\begin{gather*}
|1-F_i(z)|\leq |z|^{i+1}\left(
\sum\limits_{k=i+1}
^\infty|a_k|\right)=|z|^{i+1},
\end{gather*}
whenever $|z|\leq 1$.
Since $|z|=\left|\frac{\lambda_\ast-\lambda_{2i}}{\lambda_\ast-\lambda}\right|\leq \frac{\varepsilon}{D}< 1$ for
suf\/f\/iciently large $i\in\mathbb N$, we have
\begin{gather*}
\sum\limits_{i=1}
^\infty\left|1-\frac{\lambda-\lambda_{2i}}{\lambda_\ast-\lambda}E_i\left(\frac{\lambda_\ast-\lambda_{2i}}{\lambda_\ast-\lambda}\right)\right|\leq
\sum\limits_{i=1}
^\infty\left(\frac{\varepsilon}{D}\right)^{i+1}<\infty.
\end{gather*}

We proved that the whole series of the l.h.s.\ of~\eqref{aus} is dominated by a~uniformly convergent series,
whenever $\lambda\in K$ and $(t,x)\in\mathbb R^2$.
By standard facts concerning inf\/inite products, the theorem is proved.
\end{proof}

Now, we prove the following
\begin{lemma}
Let $K \subset G$ be a~compact subset, say $K=\{\lambda\in G\,|\,|\lambda|< l\}$.
Then the functions $x\mapsto U_{g,x}(t,x,\lambda)$ are uniformly bounded in $\mathbb R$, for $\lambda\in K$ and
uniformly in $t\in\mathbb R$.
Hence the maps $x\mapsto U_g(t,x,\lambda)$ converge uniformly on compact subsets of $\mathbb R$ to $U(t,x,\lambda)$ for
$\lambda\in K$ and $t\in\mathbb R$.
\end{lemma}

\begin{proof}
The derivative $U_{g,x}(t,x,\lambda)$ can be estimated as follows:
\begin{gather*}
|U_{g,x}|\leq \alpha|U_g|+\frac{2p}{\mathcal M}\frac{k_g(0^+)}{\prod\limits_{j=1}^gP_j^{(g)}}
\left(\sum\limits_{j=1}^g\frac{|l P_{j,x}^{(g)}|}{(P_j^{(g)})^2}\prod\limits_{k\neq j}
\left(\frac{\lambda-P_k^{(g)}}{\lambda-\lambda_\ast}\right)E_k\left(\frac{\lambda_\ast-\lambda_{2k}}{\lambda_\ast-\lambda}\right)\right),
\end{gather*}
where $\alpha= {\sup\limits_{x\in\mathbb R}}\left|\frac{p_x}{p}-\frac{\mathcal M_x}{\mathcal M}\right|$.
Since $U_g$ is uniformly bounded, the only thing to check is that
\begin{gather*}
\sum\limits_{j=1}^\infty\frac{|P_{j,x}^{(g)}|}{(P_j^{(g)})^2}
\end{gather*}
converges, because
\begin{gather*}
\left|\prod\limits_{k\neq j}\left(\frac{\lambda-P_k^{(g)}}{\lambda-\lambda_\ast}\right)
E_k\left(\frac{\lambda_\ast-\lambda_{2k}}{\lambda_\ast-\lambda}\right)\right|
\end{gather*}
is bounded uniformly with respect to $g\in\mathbb N$ and $(t,x)\in\mathbb R^2$, whenever $\lambda\in K$.

It can be shown (see~\cite{JZ3,JZ5,JZ6}) that
\begin{gather*}
\big|P_{j,x}^{(g)}(t,x)\big|\leq C\sqrt{d_j},
\end{gather*}
where $C$ does not depend on $g\in\mathbb R$ and $(t,x)\in\mathbb R^2$, hence the series
\begin{gather*}
\sum\limits_{j=1}^\infty\frac{|P_{j,x}^{(g)}|}{(P_j^{(g)})^2}\leq C
\sum\limits_{j=1}^\infty\frac{\sqrt{d_j}}{h_{0j}^2}<\infty.
\end{gather*}
The lemma is proved.
\end{proof}

Def\/ine the matrix $B_g(t,x,\lambda)$ as in Section~\ref{section3}.
This matrix has entries depending on $U_g$ and two more functions $T_g$ and $V_g$ def\/ined so as to satisfy~\eqref{TG}
and~\eqref{VG} respectively.
Moreover the equation~\eqref{det} and the stationary zero-curvature relation $-B_{g,x}+[A,B_g]=0$ hold.

Since $\{U_g\}$ and $\{U_{g,x}\}$ are uniformly bounded on every compact subset of $\mathbb R$, it follows that~$\{T_g\}$ is uniformly bounded on every compact subset of $\mathbb R$.
Using~\eqref{VG} and~\eqref{det}, one easily shows that $\{V_g\}$ is uniformly bounded on each compact subset of $\mathbb R$ as well.
This tells us that: (1) the matrices $B_g$ converge to a~matrix $B$ in every compact subset $K \subset G$
and uniformly on compact subsets of $\mathbb R^2$; (2) writing the stationary zero-curvature relation as
\begin{gather*}
B_g(t,x,\lambda)-B_g(t,0,\lambda)=\int_0^x[A_g(t,s,\lambda),B_g(t,s,\lambda)]ds,
\end{gather*}
we can use the bounded convergence theorem to conclude that also
\begin{gather*}
B(t,x,\lambda)-B(t,0,\lambda)=\int_0^x[A(t,s,\lambda),B(t,s,\lambda)]ds,
\end{gather*}
i.e.,
\begin{gather*}
-B_x(t,x,\lambda)+[A(t,x,\lambda),B(t,x,\lambda)]=0.
\end{gather*}

Now we specify the $t$-dependence of $a(t,x)=(p(t,x),q(t,x),y(t,x))$, as follows.
Fix a~number $r\in\mathbb N$, and introduce a~matrix $B_r$ as in Section~\ref{section3}.
Again, we introduce the polynomial
\begin{gather*}
U_r(t,x,\lambda)=\sum\limits_{j=0}^rp(t,x)f_j(t,x)\lambda^j.
\end{gather*}
In correspondence with every $0\leq r<g$, there exists a~matrix $B_r^{(g)}(t,x,\lambda)$, together with a~polynomial
$U_r^{(g)}$ and functions $T_r^{(g)}$ and $V_r^{(g)}$ satisfying the system~\eqref{h}.
The polynomial $U_r^{(g)}$ can be found recursively using the relation~\eqref{recur}.
The corresponding poles $P_1^{(g)}(t,x),\dots,P_g^{(g)}(t,x)$ solve the system
\begin{gather*}
P_{i,x}^{(g)}(t,x)=\frac{-\mathcal M(t,x)k_g(P_i^{(g)}(t,x))
\prod\limits_{i=1}^gP_i^{(g)}(t,x)}{p(t,x)k_g(0^+)\prod\limits_{j\neq i}(P_j^{(g)}(t,x)-P_i^{(g)}(t,x))}
\qquad
(\text{as in~\eqref{pole}}),
\\
P_{i,t}^{(g)}(t,x)=\frac{U_r^{(g)}(t,x,P_i^{(g)}(t,x))}{(P_i^{(g)}(t,x))^k}P^{(g)}_{i,x}(t,x).
\end{gather*}

A direct analysis shows that each coef\/f\/icient of the polynomial $U_r^{(g)}$ is well def\/ined, since it is a~linear
combination of at most $r$ symmetric functions of the poles $P_1^{(g)}(t,x),\dots,P_g^{(g)}(t,x)$, plus possibly
a~uniformly bounded function.
To make this clearer, we write down the form of the f\/irst of these coef\/f\/icients when $k=0$ (the case when $k\neq 0$ is
similar).
For f\/ixed $t\in\mathbb R$, set
\begin{gather*}
H_g(t,x)=\frac{\mathcal M(t,0)}{\mathcal M(t,x)}\prod\limits_{i=1}^g\frac{P_i^{(g)}(t,0)}{P_i^{(g)}(t,x)}.
\end{gather*}
We have
\begin{gather*}
f_r^{(g)}= {c_r^{(g)}H_g(t,x)},
\\
f_{r-1}^{(g)}= {H_g(t,x)\left[c_{r-1}^{(g)}+c_r^{(g)}\left(\sum\limits_{i=1}^g P_i^{(g)}(t,x)-P_i^{(g)}(t,0)\right)\right]},
\\
f_{r-2}^{(g)}=H_g(t,x)\Bigg[c_{r-2}^{(g)}-c_{r-1}^{(g)}\left(\sum\limits_{i=1}^g(P_i^{(g)}(t,x)-P_i^{(g)}(t,0))\right)
\\
\phantom{f_{r-2}^{(g)}=}
+c_r^{(g)}\Bigg(\sum\limits_{i<j}^g(P_i^{(g)}(t,x)-P_i^{(g)}(t,0))(P_j^{(g)}(t,x)-P_j^{(g)}(t,0))\Bigg) \Bigg],
\\
\cdots\cdots \cdots\cdots\cdots\cdots\cdots\cdots\cdots\cdots\cdots\cdots\cdots\cdots\cdots\cdots\cdots\cdots\cdots\cdots\cdots\cdots
\end{gather*}

These formulas imply that the coef\/f\/icients $f_j^{(g)}(t,x)$ converge, as $g\rightarrow\infty$, to coef\/f\/icients
$f_j(t,x)$ uniformly on compact subsets of $\mathbb R^2$ ($1\leq j\leq r$), hence a~polynomial $U_r(t,x,\lambda)$ is
well def\/ined as the limit, as $g\rightarrow\infty$, of the polynomials $U_r^{(g)}(t,x,\lambda)$.
Using the same arguments as those applied in the discussion of $B(t,x,\lambda)$, there is a~well def\/ined matrix
$B_r(t,x,\lambda)$ which can be obtained as the limit, as $g\rightarrow\infty$, of the matrices
$B_r^{(g)}(t,x,\lambda)$.
Using the bounded convergence theorem again, we conclude that the zero-curvature relation
\begin{gather*}
A_t(t,x,\lambda)-B_{r,x}(t,x,\lambda)+[A(t,x,\lambda),B_r(t,x,\lambda)]=0
\end{gather*}
holds, i.e., the Sturm--Liouville hierarchy is solved, and has solutions $q(t,x)$ and $y(t,x)$ such that
\begin{gather*}
y(t,x)=\frac{\mathcal M^2(t,x)}{4p(t,x)\lambda_0}\prod\limits_{k\in\mathbb N}\frac{P_k^2(t,x)}{\lambda_{2k-1}\lambda_{2k}},
\\
q(t,x)=y(t,x)\left(\lambda_0+\sum\limits_{k\in\mathbb N}\lambda_{2k-1}+\lambda_{2k}-2P_k(t,x)\right)+\tilde q(t,x),
\end{gather*}
where
\begin{gather*}
\tilde q(t,x)=-\left(\frac{(p(t,x)y(t,x))_x}{4y(t,x)}\right)_x+\left(\frac{(p(t,x)y(t,x))_x}{4y(t,x)}\right)^2.
\end{gather*}
The poles $P_1(t,x),\dots, P_g(t,x),\dots$ move according to
\begin{gather*}
P_{j,x}(t,x)=\pm\frac{\mathcal M(t,x)}{\sqrt{\lambda_0}p(t,x)}\left(\prod\limits_{k\in\mathbb N}
\frac{P_k(t,x)}{\sqrt{\lambda_{2k-1}\lambda_{2k}}}\right)\sqrt{(\lambda_{2j}-P_j(t,x))(P_j(t,x)-\lambda_{2j-1})}\times
\\
\phantom{P_{j,x}(t,x)=}
\times\left(\prod\limits_{k\neq j}\frac{\sqrt{(\lambda_{2k-1}-P_j(t,x))(\lambda_{2k}-P_j(t,x))}}{P_k(t,x)-P_j(t,x)}\right)\sqrt{P_j(x)-\lambda_0}
\end{gather*}
and
\begin{gather*}
P_{j,t}(t,x)=\frac{U_r(t,x,P_j(t,x))}{P_j^k(t,x)}P_{j,x}(t,x),
\end{gather*}
where $U_r(t,x,\lambda)$ is def\/ined as the (pointwise) limit as $g\rightarrow\infty$ of the polynomial
$U_r^{(g)}(t,x,\lambda)$.

\pdfbookmark[1]{References}{ref}
\LastPageEnding


\begin{thebibliography}{99}
\footnotesize\itemsep=0pt

\bibitem{AF}
Alber M.S., Fedorov Y.N., Algebraic geometrical solutions for certain evolution
  equations and {H}amiltonian f\/lows on nonlinear subvarieties of generalized
  {J}acobians, \href{http://dx.doi.org/10.1088/0266-5611/17/4/329}{\textit{Inverse Problems}} \textbf{17} (2001), 1017--1042.

\bibitem{BSS}
Beals R., Sattinger D.H., Szmigielski J., Multipeakons and the classical moment
  problem, \href{http://dx.doi.org/10.1006/aima.1999.1883}{\textit{Adv. Math.}} \textbf{154} (2000), 229--257,
  \href{http://arxiv.org/abs/solv-int/9906001}{solv-int/9906001}.

\bibitem{Bel}
Belokolos E.D., Bobenko A.I., Enol'skii V.Z., Its A.R., Matveev V.B.,
  Algebro-geometric approach to nonlinear integrable equations, \textit{Springer Series
  in Nonlinear Dynamics}, Springer-Verlag, Berlin, 1994.

\bibitem{BM}
Boutet~de Monvel A., Egorova I., On solutions of nonlinear {S}chr\"odinger
  equations with {C}antor-type spectrum, \href{http://dx.doi.org/10.1007/BF02843151}{\textit{J.~Anal. Math.}} \textbf{72}
  (1997), 1--20.

\bibitem{CH}
Camassa R., Holm D.D., An integrable shallow water equation with peaked
  solitons, \href{http://dx.doi.org/10.1103/PhysRevLett.71.1661}{\textit{Phys. Rev. Lett.}} \textbf{71} (1993), 1661--1664,
  \href{http://arxiv.org/abs/patt-sol/9305002}{patt-sol/9305002}.

\bibitem{Cr}
Craig W., The trace formula for {S}chr\"odinger operators on the line,
  \href{http://dx.doi.org/10.1007/BF02125131}{\textit{Comm. Math. Phys.}} \textbf{126} (1989), 379--407.

\bibitem{DMN}
Dubrovin B.A., Matveev V.B., Novikov S.P., Nonlinear equations of
  {K}orteweg--de {V}ries type, f\/inite-band linear operators and {A}belian
  varieties, \href{http://dx.doi.org/10.1070/RM1976v031n01ABEH001446}{\textit{Russ. Math. Surv.}} \textbf{31} (1976), 59--146.

\bibitem{Ego1}
Egorova I.E., On a class of almost periodic solutions of the {K}d{V} equation
  with a nowhere dense spectrum, \textit{Russian Acad. Sci. Dokl. Math.}
  \textbf{45} (1993), 290--293.

\bibitem{Ego}
Egorova I.E., The {C}auchy problem for the {K}d{V} equation with almost
  periodic initial data whose spectrum is nowhere dense, in Spectral operator
  theory and related topics, \textit{Adv. Soviet Math.}, Vol.~19, Amer. Math.
  Soc., Providence, RI, 1994, 181--208.

\bibitem{FJZ}
Fabbri R., Johnson R., Zampogni L., Nonautonomous dif\/ferential systems in two
  dimensions, in Handbook of Dif\/ferential Equations: Ordinary Dif\/ferential
  Equations, {V}ol.~{IV}, \href{http://dx.doi.org/10.1016/S1874-5725(08)80007-7}{\textit{Handb. Differ. Equ.}}, Elsevier/North-Holland,
  Amsterdam, 2008, 133--268.

\bibitem{GGKM}
Gardner C.S., Greene J.M., Kruskal M.D., Miura R.M., Method for solving the
  Korteweg--de Vries equation, \href{http://dx.doi.org/10.1103/PhysRevLett.19.1095}{\textit{Phys. Rev. Lett.}} \textbf{19} (1967),
  1095--1097.

\bibitem{GH}
Gesztesy F., Holden H., Algebro-geometric solutions of the {C}amassa--{H}olm
  hierarchy, \href{http://dx.doi.org/10.4171/RMI/339}{\textit{Rev. Mat. Iberoam.}} \textbf{19} (2003), 73--142,
  \href{http://arxiv.org/abs/nlin.SI/0105021}{nlin.SI/0105021}.

\bibitem{GH1}
Gesztesy F., Holden H., Soliton equations and their algebro-geometric
  solutions. {V}ol.~{I}. $(1+1)$-dimensional continuous models,
  \href{http://dx.doi.org/10.1017/CBO9780511546723}{\textit{Cambridge Studies in Advanced Mathematics}}, Vol.~79, Cambridge
  University Press, Cambridge, 2003.

\bibitem{GWZ}
Gesztesy F., Karwowski W., Zhao Z., Limits of soliton solutions, \href{http://dx.doi.org/10.1215/S0012-7094-92-06805-0}{\textit{Duke
  Math.~J.}} \textbf{68} (1992), 101--150.

\bibitem{J86b}
Johnson R., Exponential dichotomy, rotation number, and linear dif\/ferential
  operators with bounded coef\/f\/icients, \href{http://dx.doi.org/10.1016/0022-0396(86)90125-7}{\textit{J.~Differential Equations}}
  \textbf{61} (1986), 54--78.

\bibitem{J86}
Johnson R., On the {S}ato--{S}egal--{W}ilson solutions of the {K}-d{V}
  equation, \href{http://dx.doi.org/10.2140/pjm.1988.132.343}{\textit{Pacific~J. Math.}} \textbf{132} (1988), 343--355.

\bibitem{JM}
Johnson R., Moser J., The rotation number for almost periodic potentials,
  \href{http://dx.doi.org/10.1007/BF01208484}{\textit{Comm. Math. Phys.}} \textbf{84} (1982), 403--438.



\bibitem{JZ1}
Johnson R., Zampogni L., On the inverse {S}turm--{L}iouville problem,
  \href{http://dx.doi.org/10.3934/dcds.2007.18.405}{\textit{Discrete Contin. Dyn. Syst.}} \textbf{18} (2007), 405--428.

\bibitem{JZ2}
Johnson R., Zampogni L., Description of the algebro-geometric
  {S}turm--{L}iouville coef\/f\/icients, \href{http://dx.doi.org/10.1016/j.jde.2007.09.013}{\textit{J.~Differential Equations}}
  \textbf{244} (2008), 716--740.

\bibitem{JZ3}
Johnson R., Zampogni L., Some remarks concerning ref\/lectionless
  {S}turm--{L}iouville potentials, \href{http://dx.doi.org/10.1142/S0219493708002391}{\textit{Stoch. Dyn.}} \textbf{8} (2008),
  413--449.

\bibitem{JZ4bis}
Johnson R., Zampogni L., On the {C}amassa--{H}olm and {K}-d{V} hierarchies,
  \href{http://dx.doi.org/10.1007/s10884-010-9167-0}{\textit{J.~Dynam. Differential Equations}} \textbf{22} (2010), 331--366.

\bibitem{JZ4}
Johnson R., Zampogni L., Remarks on a paper of {K}otani concerning generalized
  ref\/lectionless {S}chr\"odinger potentials, \href{http://dx.doi.org/10.3934/dcdsb.2010.14.559}{\textit{Discrete Contin. Dyn.
  Syst. Ser.~B}} \textbf{14} (2010), 559--586.

\bibitem{JZ5}
Johnson R., Zampogni L., The {S}turm--{L}iouville hierarchy of evolution
  equations, \textit{Adv. Nonlinear Stud.} \textbf{11} (2011), 555--591.

\bibitem{JZ6}
Johnson R., Zampogni L., The {S}turm--{L}iouville hierarchy of evolution
  equations.~{II}, \textit{Adv. Nonlinear Stud.} \textbf{12} (2012), 501--532.

\bibitem{JZ7}
Johnson R., Zampogni L., On a Gel'fand--Levitan theory for the Sturm--Liouville
  operator, {P}reprint.

\bibitem{Kn}
Knopp K., Funktionentheorie. II.~Anwendungen und Weiterf\"uhrung der
  allgemeinen Theorie, de Gruyter, Berlin, 1981.

\bibitem{Ko}
Kotani S., Kd{V} f\/low on generalized ref\/lectionless potentials,
  \textit{J.~Math. Phys. Anal. Geometry} \textbf{4} (2008), 490--528.

\bibitem{LL1}
Lax P.D., Levermore C.D., The small dispersion limit of the {K}orteweg--de
  {V}ries equation.~{I}, \href{http://dx.doi.org/10.1002/cpa.3160360302}{\textit{Comm. Pure Appl. Math.}} \textbf{36} (1983),
  253--290.

\bibitem{LL2}
Lax P.D., Levermore C.D., The small dispersion limit of the {K}orteweg--de
  {V}ries equation.~{II}, \href{http://dx.doi.org/10.1002/cpa.3160360503}{\textit{Comm. Pure Appl. Math.}} \textbf{36} (1983),
  571--593.

\bibitem{LL3}
Lax P.D., Levermore C.D., The small dispersion limit of the {K}orteweg--de
  {V}ries equation.~{III}, \href{http://dx.doi.org/10.1002/cpa.3160360606}{\textit{Comm. Pure Appl. Math.}} \textbf{36} (1983),
  809--829.

\bibitem{Le}
Levitan B.M., Approximation of inf\/inite-zone by f\/inite-zone potentials,
  \href{http://dx.doi.org/10.1070/IM1983v020n01ABEH001339}{\textit{Math. USSR Izv.}} \textbf{20} (1983), 55--87.

\bibitem{Levitan}
Levitan B.M., On the closure of the set of f\/inite-zone potentials,
  \href{http://dx.doi.org/10.1070/SM1985v051n01ABEH002847}{\textit{Math. USSR Sb.}} \textbf{51} (1985), 67--89.

\bibitem{Le1}
Levitan B.M., Inverse {S}turm--{L}iouville problems, VSP, Zeist, 1987.

\bibitem{Lu}
Lundina D., Compactness of the set of ref\/lectionless potentials, \textit{Teor.
  Funkts. Funkts. Anal. Prilozh.} \textbf{44} (1985), 55--66.

\bibitem{Ma}
Marchenko V.A., Sturm--{L}iouville operators and applications, \textit{Operator
  Theory: Advances and Applications}, Vol.~22, Birkh\"auser Verlag, Basel,
  1986.

\bibitem{Ma1}
Marchenko V.A., The {C}auchy problem for the {K}d{V} equation with
  nondecreasing initial data, in What is Integrability?, \textit{Springer Ser.
  Nonlinear Dynam.}, Springer, Berlin, 1991, 273--318.

\bibitem{Mar}
Marchenko V.A., Ostrovsky I.V., Approximation of periodic by f\/inite-zone
  potentials, \textit{Sel. Math. Sov.} \textbf{6} (1987), 101--136.

\bibitem{MK}
McKean H.P., The {L}iouville correspondence between the {K}orteweg--de {V}ries
  and the {C}amassa--{H}olm hierar\-chies, \href{http://dx.doi.org/10.1002/cpa.10083}{\textit{Comm. Pure Appl. Math.}}
  \textbf{56} (2003), 998--1015.

\bibitem{Mk1}
McKean H.P., van Moerbeke P., The spectrum of {H}ill's equation,
  \href{http://dx.doi.org/10.1007/BF01425567}{\textit{Invent. Math.}} \textbf{30} (1975), 217--274.

\bibitem{NMPZ}
Novikov S., Manakov S.V., Pitaevski{\u\i} L.P., Zakharov V.E., Theory of
  solitons. The inverse scattering method, Contemporary Soviet Mathematics,
  Consultants Bureau, New York, 1984.

\bibitem{Sa}
Sato M., Soliton equations as dynamical systems on an inf\/inite dimensional
  Grassmann manifolds, in Random Systems and Dynamical Systems (Kyoto, 1981),
  \textit{RIMS Kokyuroku}, Vol. 439, Kyoto, 1981, 30--46.

\bibitem{SW}
Segal G., Wilson G., Loop groups and equations of {K}d{V} type, \textit{Inst.
  Hautes \'Etudes Sci. Publ. Math.}  (1985), 5--65.

\bibitem{Ven}
Venakides S., The {K}orteweg--de {V}ries equation with small dispersion: higher
  order {L}ax--{L}evermore theory, \href{http://dx.doi.org/10.1002/cpa.3160430303}{\textit{Comm. Pure Appl. Math.}} \textbf{43}
  (1990), 335--361.

\bibitem{Z1}
Zampogni L., On algebro-geometric solutions of the {C}amassa--{H}olm hierarchy,
  \textit{Adv. Nonlinear Stud.} \textbf{7} (2007), 345--380.

\bibitem{Z2}
Zampogni L., On inf\/inite order {K}-d{V} hierarchies, \textit{J.~Appl. Funct. Anal.} \textbf{4} (2009), 140--170.

\end{thebibliography}
\end{document}